\documentclass{aa}  
\usepackage{natbib,twoopt}
\usepackage[breaklinks=true]{hyperref} 
\bibpunct{(}{)}{;}{a}{}{,}             
\makeatletter
  \newcommandtwoopt{\citeads}[3][][]{\href{http://adsabs.harvard.edu/abs/#3}%
    {\def\hyper@linkstart##1##2{}%
     \let\hyper@linkend\@empty\citealp[#1][#2]{#3}}}
  \newcommandtwoopt{\citepads}[3][][]{\href{http://adsabs.harvard.edu/abs/#3}%
    {\def\hyper@linkstart##1##2{}%
     \let\hyper@linkend\@empty\citep[#1][#2]{#3}}}
  \newcommandtwoopt{\citetads}[3][][]{\href{http://adsabs.harvard.edu/abs/#3}%
    {\def\hyper@linkstart##1##2{}%
     \let\hyper@linkend\@empty\citet[#1][#2]{#3}}}
  \newcommandtwoopt{\citeyearads}[3][][]%
    {\href{http://adsabs.harvard.edu/abs/#3}
    {\def\hyper@linkstart##1##2{}%
     \let\hyper@linkend\@empty\citeyear[#1][#2]{#3}}}
\makeatother
\usepackage{graphicx}
\usepackage{txfonts}
\usepackage{ulem}
\usepackage{color}

\newcommand{\kms}{km\ s$^{-1}$}
\newcommand{\ms}{m\ s$^{-1}$}
\newcommand{\HTWOO}{$\textrm{H}_2 \textrm{O}$ }

\newcommand{\tauboo}{$\tau$ Boo\, }
\newcommand{\he}{He {\sc i} }

\begin{document} 

   \title{A search for \he airglow emission from the hot Jupiter $\tau$ Boo b}

    \titlerunning{\he airglow emission from $\tau$ Boo b}

 \author{Yapeng Zhang \inst{1}, I.A.G. Snellen \inst{1},  P. Molli\`ere \inst{1,2}, F. J. Alonso-Floriano \inst{1}, R. K. Webb \inst{3}, M. Brogi \inst{3, 4, 5}, A. Wyttenbach \inst{1,6}}
	
   \institute{1: Leiden Observatory, Leiden University, Postbus 9513, 2300 RA, Leiden, The Netherlands \\
   2: Max-Planck-Institut f\"{u}r Astronomie , K\"{o}nigstuhl 17, 69117 Heidelberg, Germany\\
   3: Department of Physics, University of Warwick, Coventry CV4 7AL, UK \\
   4: INAF - Osservatorio Astrofisico di Torino, Via Osservatorio 20, I-10025 Pino Torinese, Italy \\
   5: Centre for Exoplanets and Habitability, University of Warwick, Gibbet Hill, Coventry, CV4 7AL, UK \\
   6: Université Grenoble Alpes, CNRS, IPAG, 38000 Grenoble, France
             }
    \authorrunning{Yapeng Zhang et al.}
  \date{Received 13 May 2020; accepted 24 July 2020}

  \abstract
   {The helium absorption line at 10830 \AA\, originating from the metastable triplet state 2$^3$S, has been suggested as an excellent probe for the extended atmospheres of hot Jupiters and their hydrodynamic escape processes, and has recently been detected in the transmission spectra of a handful of planets. The isotropic re-emission will lead to helium airglow that may be observable at other orbital phases.}
   {The goal of this paper is to investigate the detectability of \he emission at 10830 \AA\, in the atmospheres of exoplanets using high-resolution spectroscopy, providing insights into the properties of the upper atmospheres of close-in gas giants.}
   {We estimate the expected strength of \he emission in hot Jupiters
   based on their transmission signal. We search for the \he 10830 \AA\ emission feature in $\tau$ Boo b in three nights of high-resolution spectra taken by CARMENES at the 3.5m Calar Alto telescope. The spectra in each night were corrected for telluric absorption, sky emission lines and stellar features, and shifted to the planetary rest frame to search for the emission.}
   {The \he emission is not detected in $\tau$ Boo b, reaching a 5$\sigma$ contrast limit of $4 \times 10^{-4}$ for emission line widths of $>$20 \kms. This is roughly a factor $\sim$8 above the expected level of emission (assuming a typical \he transit absorption of 1\% for hot Jupiters). This suggests that targeting the \he emission with well-designed observations using upcoming instruments such as VLT/CRIRES+ and E-ELT/HIRES is possible.
  }
   {}

    \keywords{planets and satellites: atmospheres -- 
    planets and satellites: individual ($\tau$ Boo b) -- 
    techniques: spectroscopic  -- 
    Infrared: planetary systems}
   \maketitle


\section{Introduction}
The atmospheres of close-in exoplanets are strongly affected by the high-energy irradiation from their host stars, making their upper atmospheres weakly bound and susceptible to hydrodynamic escape \citep{2019AREPS..47...67O}. 
For gas-giant planets, Lyman-$\alpha$ absorption in transit measurements has been the primary probe for their extended atmospheres, e.g., HD 209458b \citep{2003Natur.422..143V}, GJ 436b \citep{2014ApJ...786..132K, 2015Natur.522..459E} and GJ 3470b \citep{2018A&A...620A.147B}. 
Such observations trace the exospheres with neutral hydrogen extending far beyond the Roche lobe radius, and cast light on the hydrodynamic escape and mass loss, which drives the evolution of the atmospheric and bulk composition of close-in exoplanets \citep{2015Natur.522..459E}.

In addition to Ly$\alpha$, helium at 10830 \AA\, was also identified to be a powerful tracer of the extended atmospheres  \citep{2000ApJ...537..916S}, which was recently reassessed by \citet{2018ApJ...855L..11O}. This helium line has several advantages over Ly$\alpha$ because it suffers less from the interstellar absorption and can be observed from the ground with high-resolution spectrographs, opening a new window into the characterization of exospheres. 

The theoretical work by \citet{2018ApJ...855L..11O} has resulted in a breakthrough in helium measurements. Excess absorption was first detected by the Hubble Space Telescope/Wide Field Camera 3 in WASP-107b \citep{2018Natur.557...68S} and independently confirmed by ground-based observations \citep{2019A&A...623A..58A, 2020AJ....159..115K} showing an absorption level of 5.54 $\pm$ 0.27\%. 
In addition, another five planets have been reported with helium signals, namely, HAT-P-11b at a level of 1.08 $\pm$ 0.05\% \citep{2018Sci...362.1384A, 2018ApJ...868L..34M}, HD 189733b at 0.88 $\pm$ 0.04\% \citep{2018A&A...620A..97S, 2020arXiv200505676G}, WASP-69b at 3.59 $\pm$ 0.19\% \citep{2018Sci...362.1388N}, HD 209458b at 0.91 $\pm$ 0.10\% \citep{2019A&A...629A.110A}, and GJ 3470b at 1.5 $\pm$ 0.3\% \citep{2020ApJ...894...97N, 2020A&A...638A..61P}. 
Upper limits in \he absorption were derived for Kelt-9b and GJ 436b at 0.33\% and 0.41\% \citep{2018Sci...362.1388N}, for WASP-12b at 59$\pm$143 ppm over a 70\AA\, band \citep{2018RNAAS...2...44K}, for GJ 1214b at 3.8\%$\pm$4.3\% \citep{2019RNAAS...3...24C}, for K2-100b with an equivalent width less than 5.7 m\AA\,\citep{2020MNRAS.495..650G}, and AU Mic b with EW of 3.7 m\AA\,\citep{2020arXiv200613243H}.

The wide range of helium absorption levels seen in the transmission spectra of hot gas giants is likely due to the variety in the stellar radiation fields. 
The helium absorption line at 10830 \AA\, originates from atoms in a metastable triplet state $2^3$S, which is mainly populated via recombination of ionized helium. The ionization of \he requires photons with wavelengths smaller than 504 \AA, i.e. X-ray and extreme-ultraviolet (EUV). Therefore, the helium absorption is expected to be enhanced for planets irradiated with higher X-ray and EUV flux \citep{2018Sci...362.1388N}. 
However, the level of EUV irradiation may not be the only determining factor. The mid-ultraviolet radiation near 2600 \AA\, can de-populate the triplet state $2^3$S via ionization. Therefore, the strength of the helium absorption has been proposed to depend on the ratio between EUV and mid-UV stellar fluxes, and K-type stars may provide the most favorable stellar environment \citep{2019ApJ...881..133O}. Hitherto, the exoplanets with detected helium absorption generally support this trend with four out of six orbiting K-type stars. More detections will help to better understand the factors that affect the \he absorption level.

As helium atoms at the triplet state $2^3$S absorb photons to reach $2^3$P state, the opposite transition will happen concurrently, with 10830 \AA\, photons re-emitted in random directions. Based on the absorption level, we estimate the amount of emission expected from the planetary atmospheres. The study of \he airglow emission allows for better understanding of the radiation fields and level populations under non‐local thermodynamic equilibrium in the upper atmospheres. In addition, it provides a way of probing the extended atmospheres of non-transiting close-in gas giants, which have not been investigated before.

In this paper, we address the detectability of \he emission in the atmospheres of close-in exoplanets. We first calculate the expected emission level (Section~\ref{sec:theory}), then perform a case study on $\tau$ Boo b (Section~\ref{sec:tauboo}) searching for helium airglow emission in 
 high-resolution spectroscopic data from CARMENES (Calar Alto high-Resolution search for M dwarfs with Exoearths with Near-infrared and optical Echelle Spectrographs) in Section~\ref{sec:data} and \ref{sec:result}. We discuss our results and future prospects in Section~\ref{sec:discuss}.


\section{Helium emission from extended atmospheres} \label{sec:theory}

\begin{figure}
   \centering
   \includegraphics[width=\hsize]{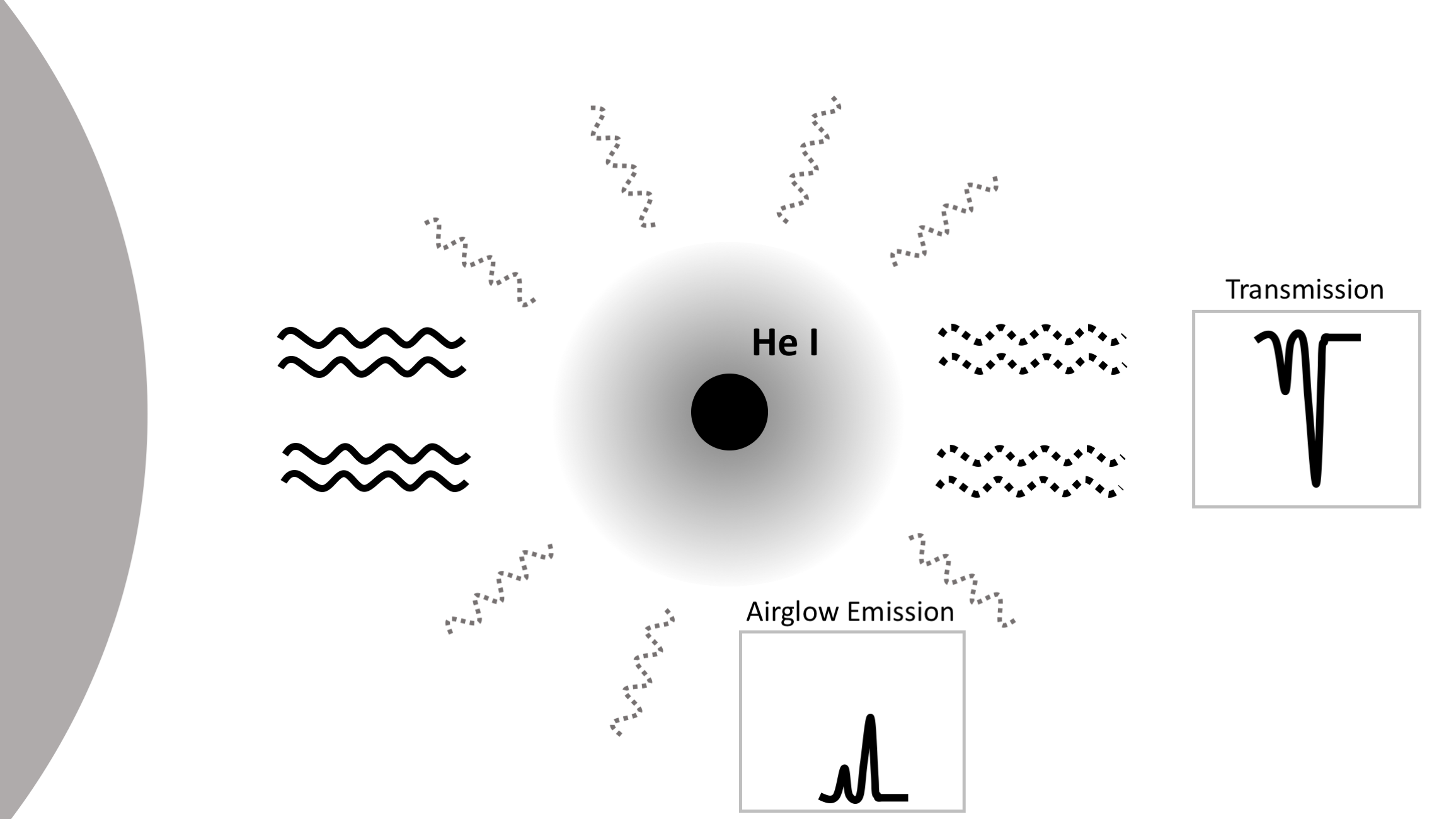}
   \caption{Illustration of \he emission from an exosphere of a close-in planet. The gray disk represents the extended atmosphere around such planet. \he absorption is observed during transit, while airglow emission can be probed at other orbital phases.}
    \label{fig:schem}%
\end{figure}

The concept of probing \he airglow emission is illustrated in Fig.~\ref{fig:schem}. 
We consider the extended atmosphere as an optically thin cloud surrounding the exoplanet. The cloud absorbs a fraction of irradiation from the host star due to the transition at 10830 \AA, resulting in the helium atoms at state $2^3$S to be excited to the higher state $2^3$P. The absorbed radiation is reemitted isotropically when the excited electrons jump back down to the metastable triplet state. Assuming the extended atmosphere has a low density and is optically thin, the reemitted photons can freely escape and be observed as an emission feature. We would like to point out that during transit the feature is seen in absorption because only a small fraction of the radiation is reemitted along the line-of-sight. Consequently, we can detect \he absorption lines during transit, while emission lines from other viewing angles.

To estimate the strength of the \he emission, the cloud is assumed to absorb a fraction of the incident stellar energy at 10830 \AA, emitting it back isotropically. Assuming the upper energy level not being (de-)populated in another way, we can estimate \citep[see][Chap.~3]{2010eapp.book.....S}
\begin{equation}
    4\pi R_\mathrm{c}^2 F_\mathrm{c}^S = F_{\ast}^S f_{\mathrm{abs}}\bigg(\frac{R_\ast}{a}\bigg)^2\pi  R_\mathrm{c}^2 
\end{equation}
where $R_\mathrm{c}$ is the radius of the cloud that absorbs radiation at that wavelength, $R_\ast$ is the stellar radius, $a$ is the orbital distance of the planet, $f_{\mathrm{abs}}$ is the fraction of energy absorbed at 10830 \AA, which is linked to the column density of helium atoms at state $2^3$S in the cloud, $F_\mathrm{c}^S$ and $F_{\ast}^S$ denote the 10830 \AA\, flux at the surface of the planet and the star respectively. 
The corresponding fluxes that we observe at Earth are 
\begin{equation} \label{eq:dillute}
   {\mathcal{F}_\mathrm{c}} = F_\mathrm{c}^S \bigg(\frac{R_\mathrm{c}}{D}\bigg)^2 , \quad  {\mathcal{F}_\ast} = F_{\ast}^S \bigg(\frac{R_\ast}{D}\bigg)^2
\end{equation}
where $D$ is the distance of the system to Earth.
Thus, the planet to star contrast at 10830 \AA\, is 
\begin{equation}\label{eq:1}
    \frac{\mathcal{F}_\mathrm{c}}{\mathcal{F}_\ast} = \frac{f_{\mathrm{abs}}R_\mathrm{c}^2}{4a^2} = \frac{R_\ast^2}{4a^2} T_\lambda = 10^{-4}\bigg(\frac{T_\lambda}{1\%}\bigg)\bigg(\frac{5}{a/R_\ast}\bigg)^2
\end{equation}
$T_\lambda$ is the excess absorption at 10830 \AA\,  that can be measured by transmission spectroscopy, with $T_\lambda = f_{\mathrm{abs}} R_\mathrm{c}^2 / R_\ast^2$.

Uncertainties regarding this estimate are further discussed in Section~\ref{sec:theory_disc}. A more detailed derivation involving radiative transfer is presented in Appendix~\ref{app}.
Given Equation~\ref{eq:1}, we expect contrast levels of $\sim 10^{-4}$, which may be reached by combining multiple nights of observations. 

\section{The $\tau$ Boo system}\label{sec:tauboo}
The hot Jupiter $\tau$ Bootis b is among the first exoplanets discovered \citep{1997ApJ...474L.115B}, orbiting a bright F7V type main-sequence star (V=4.5 mag) at a distance of 15.6 pc from Earth. 
Although the planet is not transiting, its orbital inclination was determined to be 45$^\circ$ by tracking CO absorption in its thermal dayside spectrum \citep{2012Natur.486..502B}, confirmed by measurements of H$_2$O \citep{2014ApJ...783L..29L}. 
The properties of the $\tau$ Boo system are summarized in Table \ref{tab:tauboo_info}. 
Unfortunately, because the planet is not transiting, there is no direct measure of its radius (meaning that its surface gravity is unknown), and the \he 10830 \AA\, line cannot be probed in transmission.

The system has been an interesting target for investigating star-planet interactions because of its strong stellar activity. The star has a high S-index of 0.202 \citep{2004ApJS..152..261W}, indicating strong emission cores in the Ca II H\&K lines. It exhibits an X-ray luminosity of $9\times10^{28}$ erg s$^{-1}$ \citep{1998A&AS..132..155H}, and a reconstructed EUV luminosity of $2.5\times10^{29}$ erg s$^{-1}$ \citep{2011A&A...532A...6S}, which are among the highest values found in the planet-hosting stars. The intense X-ray and EUV radiation can deposit substantial energy in the planetary atmosphere, facilitating the expansion and evaporation of gas. Although \tauboo is not a K-type star (which is possibly the most favorable spectral type for probing the helium line), however, due to the high level of stellar X-ray and EUV emission, \tauboo b could well have an extended atmosphere with a large population of helium at the triplet state, and forms a promising target to search for the \he emission at 10830 \AA.

\begin{table}
\caption{Properties of the star $\tau$ Boo (upper part) and $\tau$ Boo b (lower part).
} 
\label{tab:tauboo_info}

\begin{tabular}{lcc}
\hline
\hline
\textbf{Parameter} & \textbf{Symbol} & \textbf{Value} \\ 
\hline
Distance $\textrm{(pc)}^a$ 				& $d$ 					& $15.66 \pm 0.08$ \\
Effective temperature $(\textrm{K})^b$ 	& $T_{\textrm{eff}}$ 	& $6399 \pm 45 $ \\
Luminosity $(L_{\odot})^b$  			& $L_*$ 				& $3.06 \pm 0.16$ \\
Mass $(M_{\odot})^b$ 		   			& $M_*$ 				& $1.38 \pm 0.05$ \\
Radius $(R_{\odot})^b$ 					& $R_*$ 				& $1.42 \pm 0.08$ \\
Surface gravity $\textrm{(cgs)}^c$ 		& $\log g$ 				& $4.27^{+0.04}_{-0.02}$ \\
Systemic velocity $\textrm{(\kms)}^d$   & $\gamma$ 		& $-16.4 \pm 0.1$ \\
Rotation velocity $\textrm{(\kms)}^b$   & $v\sin (i)$           & $14.27 \pm 0.06$ \\
Semi-amplitude $\textrm{(\ms)}^e$      & $K_\ast$              &  $ 468.42 \pm 2.9$\\
Metallicity $(\textrm{dex})^{b}$		& $\left[\textrm{M}/ \textrm{H} \right]$  & $0.26 \pm 0.03$ \\
Age $\textrm{(Gyr)}^b$					& 						& $0.9 \pm 0.5$ \\ 
\hline
Orbital period $\textrm{(days)}^e$             	& $P$               & $3.312454\pm 3.3 \cdot 10^{-6}$ \\
Semi-major axis $\textrm{(AU)}^b$     		   	& $a$               & $0.049 \pm 0.003$ \\
Orbital inclination $\textrm{(deg)}^d$          & $i$               & $44.5 \pm 1.5$ \\
$\textrm{Eccentricity}^b$    	  				& $e$				& $0.011 \pm 0.006$ \\
Mass $(M_{\textrm{J}})^b$                       & $M_p$             & $6.13 \pm 0.17$ \\
Phase zero-point $\textrm{(HJD)}^e$			   	& $T_0$             & $56401.8797 \pm 0.0036$ \\
Semi-amplitude $\textrm{(\kms)}^d$              & $K_P$             &  $110.0 \pm 3.2$\\
\hline
\end{tabular}

\tablebib{ $(a)$ \citet{2018A&A...616A...1G}; $(b)$ \citet{2015A&A...578A..64B}; $(c)$ \citet{2007ApJS..168..297T}; $(d)$ \citet{2012Natur.486..502B}; $(e)$  \citet{2019A&A...625A..59J}}
\end{table}


\section{Observations and Data analysis}\label{sec:data}

We observed and analyzed the $\tau$ Boo system for five nights with the CARMENES spectrograph mounted on the 3.5m Calar-Alto Telescope \citep{2016SPIE.9908E..12Q}, on 26 March 2018, 11 May 2018, 12 March 2019, 15 March 2019 and 11 April 2019. CARMENES has two channels, VIS and NIR, covering the optical wavelength range (520-960 nm) and the near-infrared wavelength range (960-1710 nm) respectively. The resolving power is 94,600 in the VIS channel and 80,400 in the NIR channel. Each channel is fed with two fibers: fiber A targeting the star and fiber B obtaining a sky spectrum simultaneously. The \he line at 10830 \AA\, falls on echelle order 56 (10801-11001 \AA) in the NIR channel, on which we focused our analysis.

\begin{figure}
   \centering
   \includegraphics[width=\hsize]{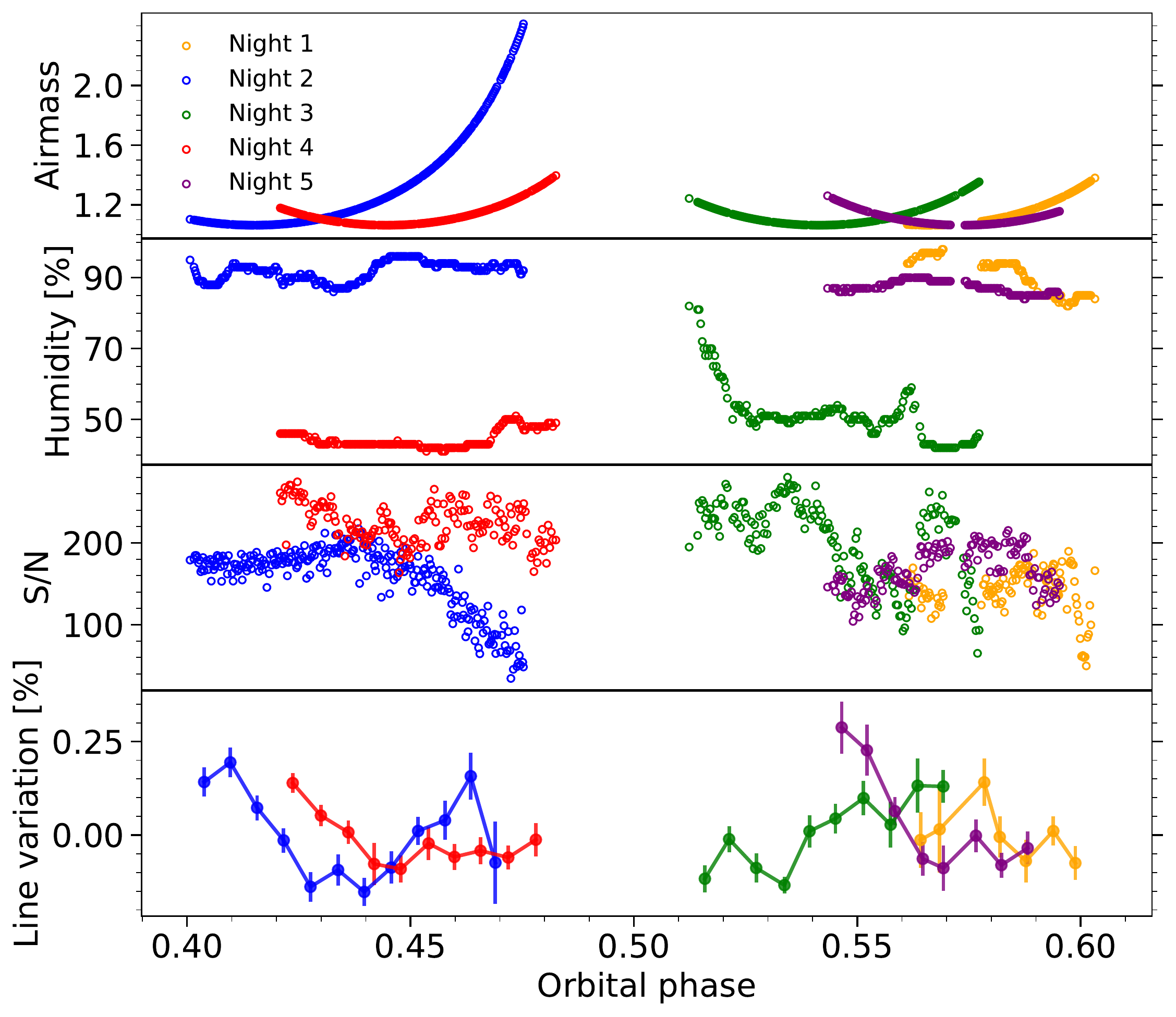}
   \caption{Variation in airmass (upper panel), relative humidity (mid panel), S/N (lower panel) and variation  around the mean of the stellar \he line (bottom panel) during observations in the different nights. The S/N of each spectrum is defined as the average signal-to-noise ratio of the continuum near the \he 10830 \AA\, line. }
    \label{fig:obs}%
\end{figure}

The details of the observations are shown in Table~\ref{tab:obs} and Fig.~\ref{fig:obs}. The observations probed thermal emission from $\tau$ Boo b, covering a wide range in planet orbital phases each night (see Fig.~\ref{fig:obs}).
The exposure times were adjusted according to the weather conditions, maintaining a S/N$\geq$100. The observations in the first two nights were taken with exposure times of 40 s with a seeing of 2$\arcsec$ and 1.4$\arcsec$ respectively, delivering lower S/N per spectrum than during the other nights with longer exposures and better weather conditions. The drop-off in S/N during the end of night 2 was due to the increasingly high airmass. The relative humidity during night 1, 2 and 5 continuously exceeded 85\%, in contrast to the lower value ($\sim$45\%) during night 4. This agrees well with the behavior of S/N in these nights. For night 3, although the relative humidity was not as high as the first two nights, it underwent strong variation overnight, possibly accounting for the drastic change in S/N. 

The standard data reduction, such as bias removal, flat fielding, cosmic ray corrections, and wavelength calibration, was performed with the CARMENES pipeline \texttt{CARACAL} v2.10 \citep{2016SPIE.9910E..0EC}. 
The output of the pipeline was provided in the observer's frame with a vacuum wavelength solution. We converted the vacuum wavelength solution into air wavelengths and used it throughout our analysis.

\begin{table*}
\centering
\caption[]{Observations Summary}
 \label{tab:obs}
\resizebox{\hsize}{!}{
\begin{tabular}{cccccccc}  
\hline\hline
\textbf{Night} & \textbf{Date} & \textbf{Proposal No.}  & \textbf{Program PI} & \textbf{No. of spectra} & \textbf{Exposure time (s)} & \textbf{On-target time (h)} & \textbf{S/N} \\ 
\hline
1 & 2018-03-26 & F18-3.5-012 & J.A.Caballero \& F.J.Alonso-Floriano & 110 & 40 & 1.2 & 140 \\
2 & 2018-05-11 & F18-3.5-012 & J.A.Caballero \& F.J.Alonso-Floriano & 261 & 40 & 2.9 & 160 \\
3 & 2019-03-12 & F19-3.5-051 & M.Brogi & 161 & 66 & 3.1 & 200 \\
4 & 2019-03-15 & F19-3.5-051 & M.Brogi  & 165 & 65 & 3.0 & 220 \\
5 & 2019-04-11 & F19-3.5-051 & M.Brogi  & 133 & 66 & 2.4 & 170 \\
\hline
\end{tabular}%
}
\end{table*}

\begin{figure}
   \centering
   \includegraphics[width=\hsize]{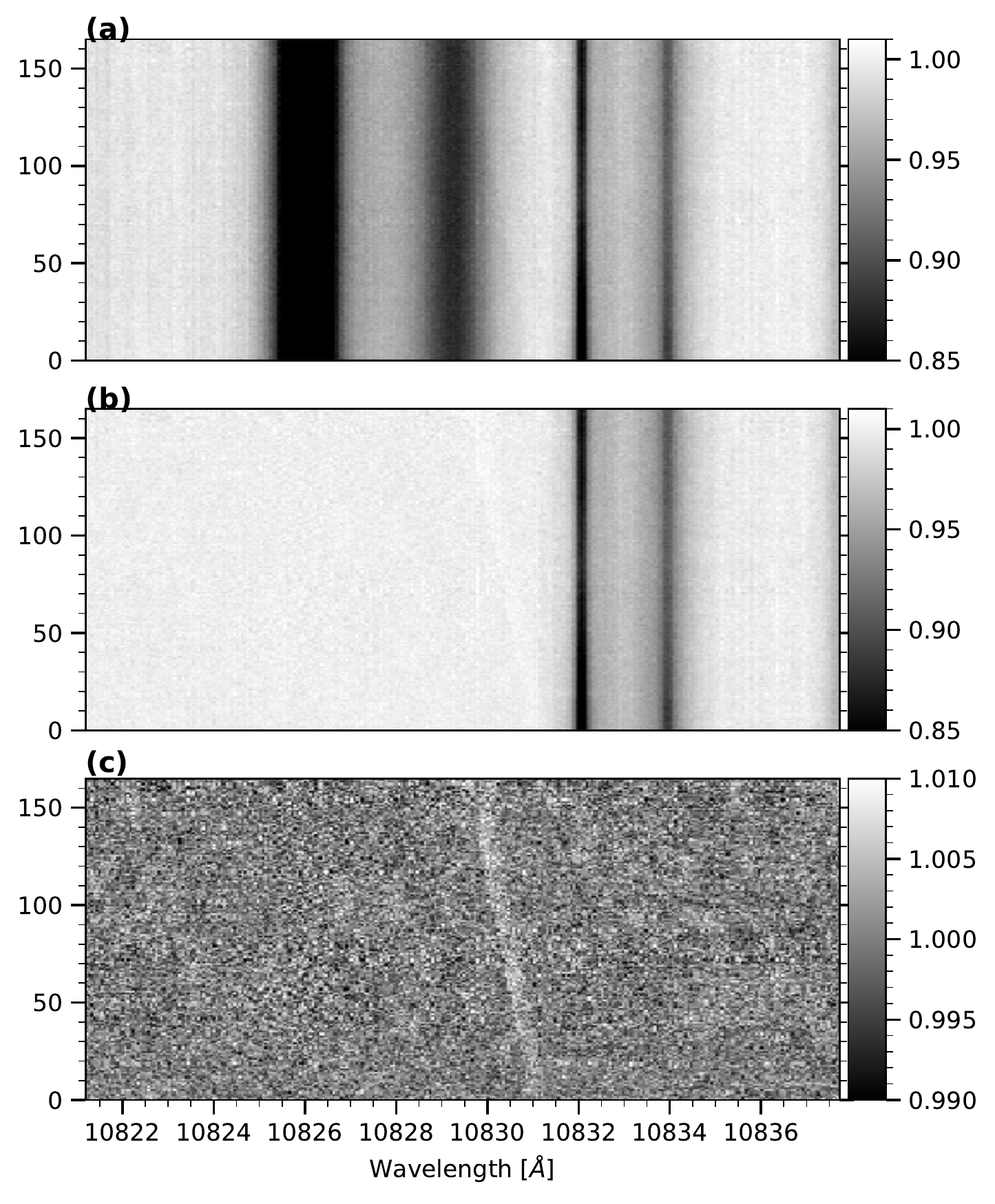}
   \caption{Illustration of our data reduction steps applied to the spectral series taken on Night 4. Each row in the matrix represents one spectrum, with the y-axis corresponding to time. As detailed in Section~\ref{sec:v1}, the steps include: (a) normalization of the continuum; (b) correction for stellar lines in each spectrum using an empirical model; (c) removal of telluric contamination by correcting the temporal variation of the flux in telluric lines using a column-by-column linear regression.
   The trail of an artificially injected planetary helium line (with a magnitude of  0.3\% and a width of 11 \kms) can be seen as a slanted white band near 10830 \AA, shifting in time due to the change in the radial component of the orbital velocity of the planet.}
    \label{fig:steps_N4}%
\end{figure}

\subsection{Data analysis}\label{sec:v1}
First, we manually corrected for additional hot pixels and cosmic rays present in the pipeline-reduced spectra by substituting those values with the linear interpolation of adjacent pixels or adjacent time series. Subsequently, each spectrum was normalized to unity using the continuum both at the blue and red side of the helium line over the ranges 10804.0-10805.2 \AA, 10818.5-10819.0 \AA\, and 10839.0-10840.3 \AA. After normalization, the spectral series of each night were handled as a two-dimensional matrix as shown in Fig.~\ref{fig:steps_N4}(a). Each row in the matrix represents one spectrum, with the frame number on the y-axis, corresponding to orbital phase or time. 

Before correcting for the telluric lines, we first removed the stellar Si line and \he triplet at 10827.1 \AA\, and 10829.09, 10830.25 and 10830.34 \AA. In order to achieve this, we built an empirical stellar model for each night by combining all spectra with S/N$>$100 in the stellar rest frame, where we set all values outside of the two stellar features to unity. This model was then shifted to the observer's rest frame, scaled and removed from each spectrum. Subsequently, the spectra were free of stellar lines in the wavelength region near the planetary helium line (see Fig.~\ref{fig:steps_N4}(b)). 

The residual matrix in Fig.~\ref{fig:steps_N4}(b) shows the telluric lines near the planetary helium signal that required removal,  including the \HTWOO absorption features at 10832.1 \AA\, and 10834.0 \AA, and the OH emission lines at 10824.7 \AA, 10829.8 \AA\, and 10831.3 \AA. 
The strengths of telluric lines vary overnight, which mainly results from the change in airmass and/or water column. 
To correct for telluric absorption lines, we measured the temporal variation of the flux at the centers of several deep \HTWOO absorption features and combined them to serve as a representation for the atmospheric change overnight. Such temporal variation was then removed using a column-by-column linear regression while avoiding the columns that contain the assumed planetary \he emission. We removed the sky emission lines in a similar way, using their mean-variation overnight. 
After telluric correction, each column of the matrix was subsequently normalized as follows. The spectra were combined via weighted average, with the weights defined as the squared S/N of each exposure, to build a master spectrum. Each individual spectrum was subsequently divided by this master spectrum to obtain the residual spectral series as shown in Fig.~\ref{fig:steps_N4}(c). 

We note that during the telluric removal and final normalisation, we masked the region in each spectrum where the planetary He line is expected to appear, so that we could avoid self-subtraction of the signal, if present. This is particularly important if the planetary \he line is broad. In the case that the line width is larger than the change of the planetary radial velocity overnight, the planetary signal would be (partially) subtracted out, making it less likely to be detected.

\subsection{Extra noise in residual spectra}

\begin{figure}
   \centering
   \includegraphics[width=\hsize]{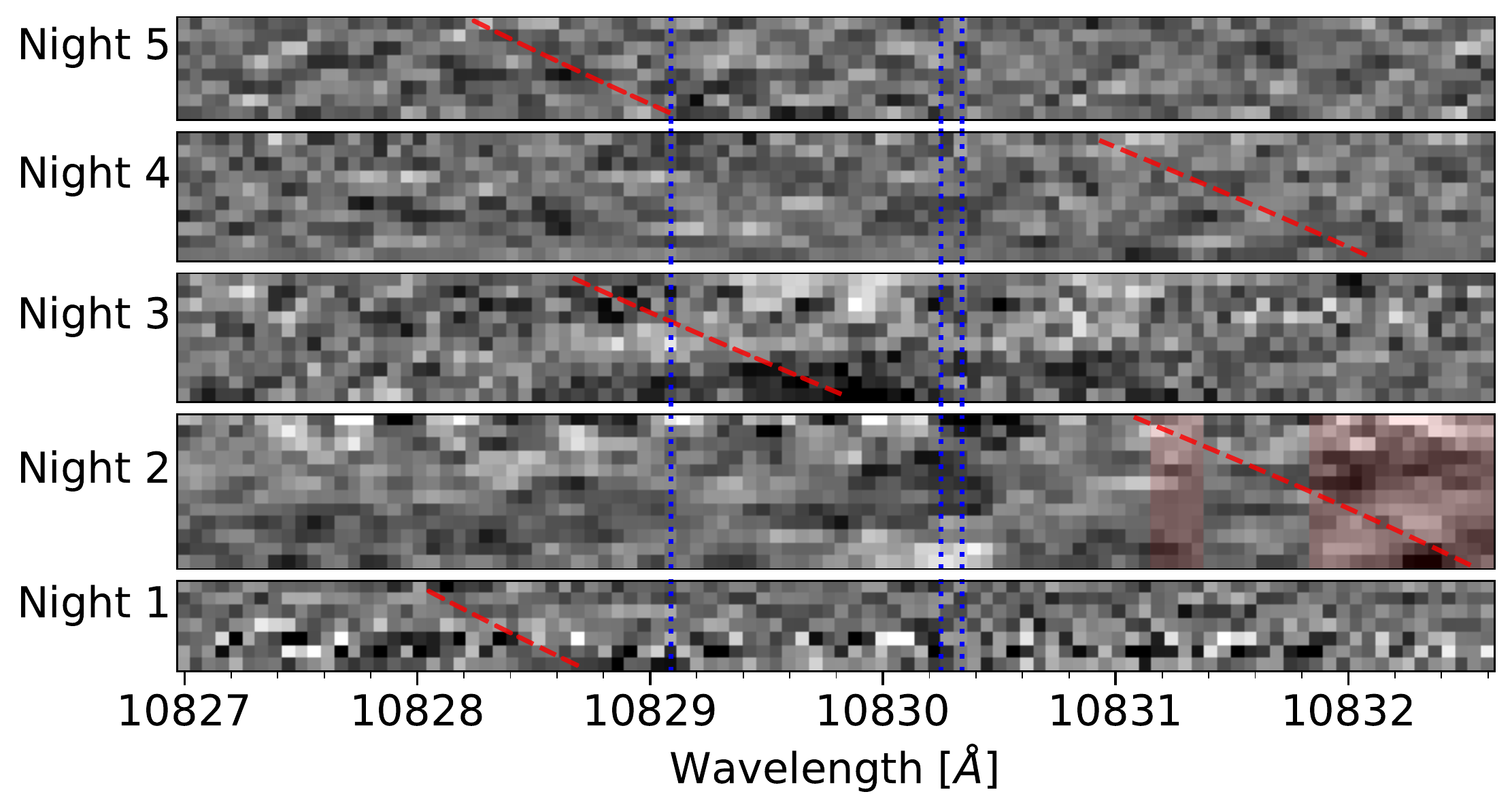}
   \caption{Residual spectral series from night 1 to 5. The y-axis represents the time or orbital phase. The residuals are binned to 0.006 in phase for clarity. The slanted dashed lines in red trace the expected planetary helium line. The dotted lines in blue denote the positions of the stellar \he triplet. The shaded region in night 2 contains the telluric \HTWOO absorption and OH emission line, overlapping with the expected planetary helium line.
   }
    \label{fig:res_bin}%
\end{figure}

After correcting for the stellar and telluric effects, we shifted the residual spectral series of all five nights into the stellar rest frame by correcting for the systemic velocity, barycentric velocity, and stellar reflex motion of $\tau$ Boo, as shown in Fig.~\ref{fig:res_bin}. The spectra were binned to 0.006 in phase (y-axis) for a better identification of the noise structure. 
We noticed that the observations in nights 2 and 3 suffer from broad noise structures, which may originate from the variability of stellar lines or the residuals of telluric corrections.

\begin{figure}
   \centering
   \includegraphics[width=\hsize]{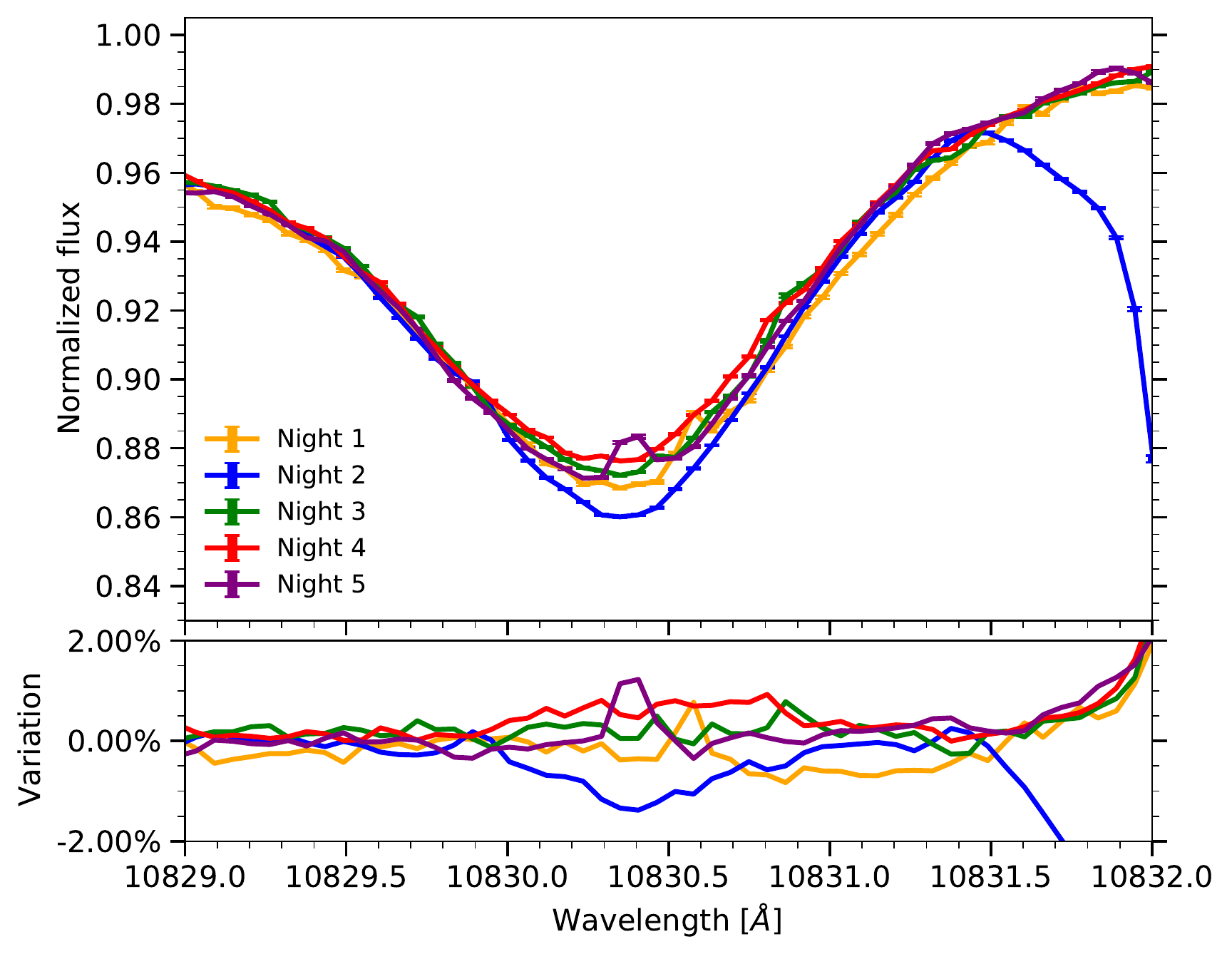}
   \caption{Upper panel: the average profile of the stellar helium line in each night. Lower panel: the difference of each profile with respect to the mean. The deep absorption line at 10832.2 \AA\,in night 2 (blue curve) is a telluric \HTWOO line. The small peaks on top of the helium line in night 1 and 5 are caused by sky OH emission.}
    \label{fig:linevar}%
\end{figure}

As we noted in section~\ref{sec:tauboo}, the star \tauboo shows a high level of chromospheric activity. The stellar \he line as a chromospheric diagnostic may undergo temporal variation due to flaring. The strength of the stellar \he line can also show periodic modulation by rotation if the line is not homogeneous on the stellar disk \citep{2017ApJ...839...97A}.
Moreover, the fast spin of \tauboo may result in residual noise features shifting in wavelength up to its rotation velocity ($\sim$15 \kms). 
To quantify the effect of stellar variability, we measured the strength of the stellar \he line at 10830 \AA\, during each night by fitting a Gaussian profile to the data and measuring the amplitude of the best-fit Gaussian profile. We define the line variation as the relative depth with respect to the average stellar spectrum of each night, and then bin the results to 0.006 in phase (see Fig.~\ref{fig:obs} bottom panel). During a single night, the variation is generally $\leq$0.25\%, which is within the noise level of the data, therefore should not significantly affect our analysis. However, it is possible that the variation can build up as we combine different frames during the night, leading to systematic noise in the combined residual spectrum. For example, the variability of \he line may be responsible for the variation in the residual flux at 10830.3 \AA\, in night 2 (see Fig.~\ref{fig:res_bin}).
We also compared the average line profile of different nights (Fig.~\ref{fig:linevar}).
On a night-to-night basis, we note that the helium line profile in night 2 is distinct from other nights, perhaps due to the influence of the nearby strong telluric line or the higher level of stellar activity. 

In addition, we note that the spectral S/N in the middle of night 3 abruptly dropped by a factor of two as shown in Fig.~\ref{fig:obs}, probably due to the changing weather condition. This may lead to the inconsistency of the data in the region from 10829.4 to 10830.2 \AA. Unfortunately, the trail of the planetary signal in night 3 resides exactly in this problematic wavelength range, overwhelmed by artifacts (see Fig~\ref{fig:res_bin}). 
As for the night 2 observations, telluric lines show a high level of strength and variability due to the high relative humidity and the variation of the precipitable water vapor, which could not completely be removed. The residuals of the deep telluric \HTWOO line at 10832.1 \AA\, and OH sky emission at 10831.3 \AA\, coincide with the trail of the planetary signal (see Fig~\ref{fig:res_bin}), making it difficult to preserve it while removing the telluric contamination. In such condition, adding the night 2 data does not enhance the S/N because more noise was introduced along with the diminished signal. Consequently, we excluded the observations of night 2 and 3 from further analysis.

Based on the ephemeris listed in Table~\ref{tab:tauboo_info}, we shifted the residual spectra to the planetary rest frame. The residual spectral series from different nights were then combined with each residual spectrum weighted by its variance over the radial velocity ranges from -150 to -50 \kms and from 50 to 150 \kms. 


\section{Result}\label{sec:result}

\begin{figure}
   \centering
   \includegraphics[width=\hsize]{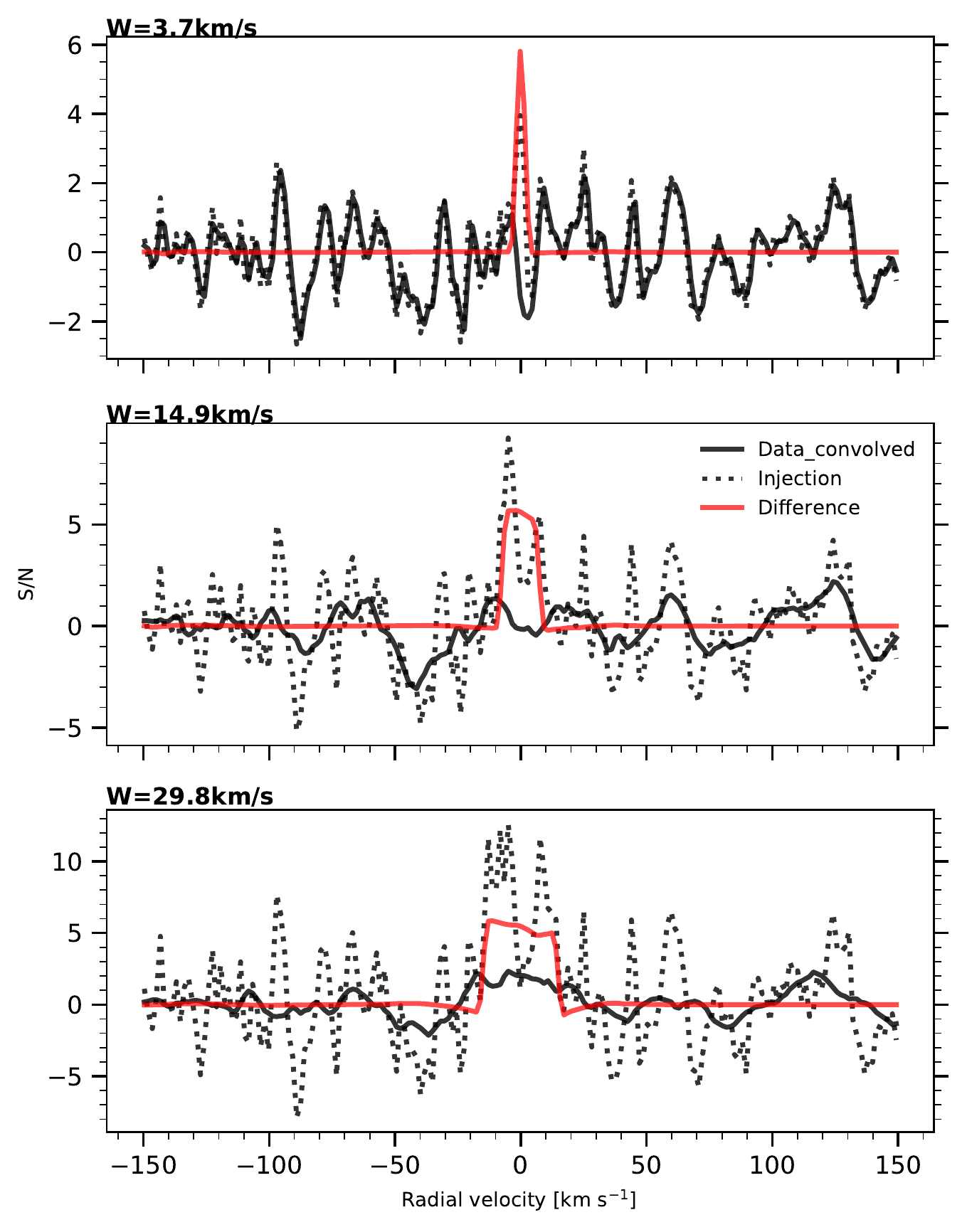}
   \caption{Combined (3 nights) residual spectra in the planetary rest frame centered at the 10830 \AA\, \he line. In the three panels, the solid black line indicates the residual spectra, boxcar-smoothed by 3.7 \kms (top), 14.9 \kms (middle), and 29.8 \kms (bottom). The dotted curves show an artificially injected signal at a S/N of 5, with the solid red curve the difference between the injected and original data. The values are scaled by the standard deviation of the observed residuals so that the y-axis represents the S/N.}
  \label{fig:res1d}%
\end{figure}

Fig.~\ref{fig:res1d} shows the time-averaged spectrum around the helium line in the planet rest-frame, with the amplitude scaled to make the standard deviation equal to unity around the targeted \he 10830 \AA\ line. We find no statistically significant signal from the planetary \he 10830 \AA\, line  within the three nights of observations. 

\subsection{Detection limits}\label{sec:limit}

\begin{figure}
   \centering
   \includegraphics[width=\hsize]{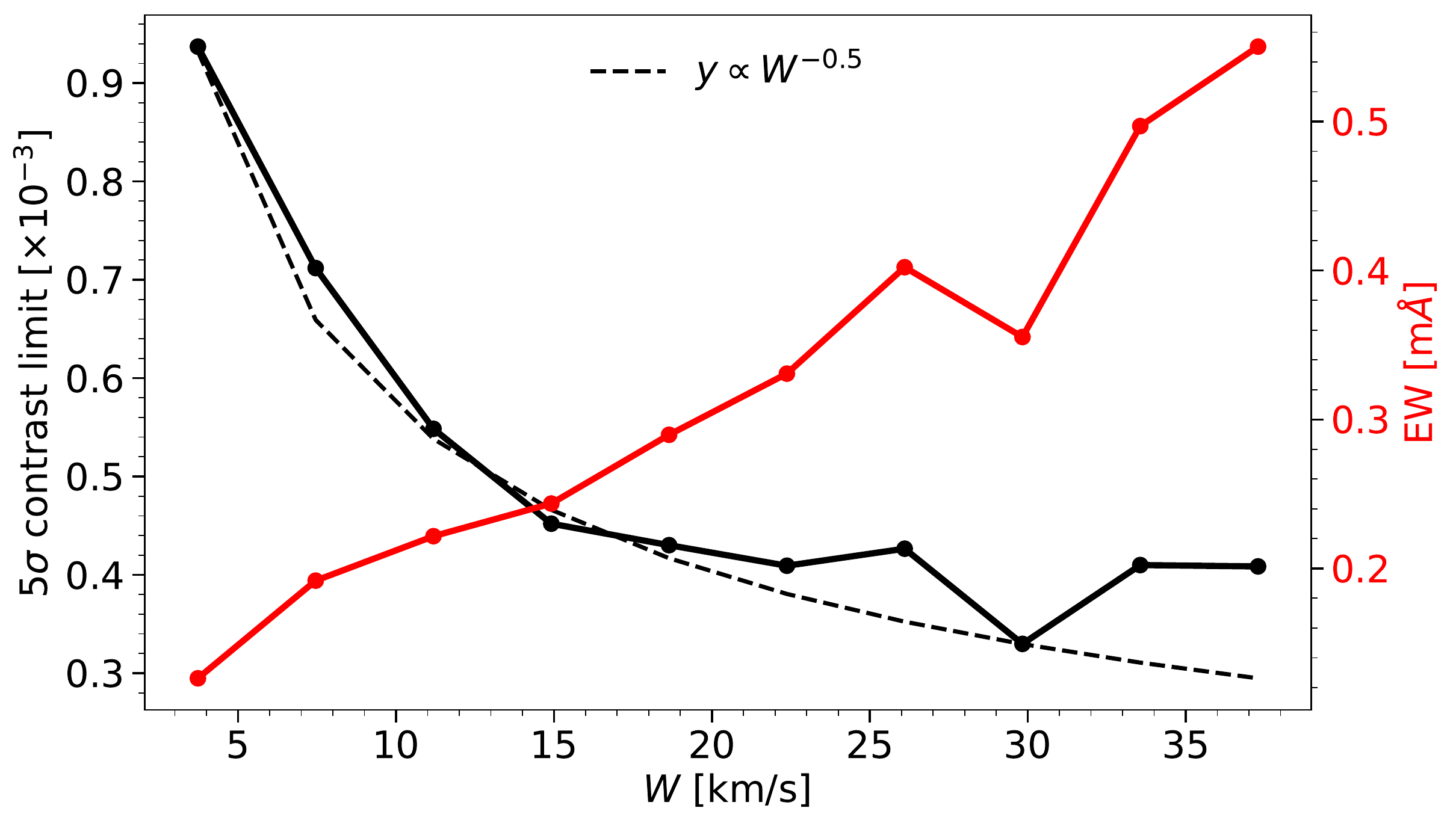}
   \caption{The $5\sigma$ detection limit of the \he airglow emission as a function of the line width of the potential signal using 3 nights of observations of \tauboo b (solid black line), and corresponding equivalent widths of the detection limit (solid red curve). The dashed line represents a scaled relation of $W^{-1/2}$, expected for pure Gaussian white noise.}
    \label{fig:limit}%
\end{figure}

As we have no information on the profile of the potential \he line originating from $\tau$ Boo b, we assumed a box profile centered at 10830.3 \AA, which has a clear definition of the line width compared to a Gaussian profile. We performed signal injections considering various widths ($W$) of the planetary emission, ranging from 3.7 \kms (equal to the instrument resolution) to 37 \kms (which is limited by the self-subtraction of broad signals). The combined residuals of both the observations and injections are shown in Fig.~\ref{fig:res1d} for three different line widths as examples. 
We regarded the original residuals (without signal injection) as the noise, and convolved it with the corresponding signal profile to take the width of the signal into account. The noise level is defined as the standard deviation (from -150 to +150 \kms) of the convolved residuals. These were then scaled by this noise level, so that the y-axis represents the signal-to-noise ratio (S/N). The strength of the signal in each injection case was determined by measuring the amplitude of the recovered signal, which is the difference between the injected and the original residuals, as shown by the red curves in Fig.~\ref{fig:res1d}. The 5$\sigma$ detection limits were thus determined as the minimum line emission leading to a recovered S/N of 5 in our injection tests. The detection limit as a function of the line width is shown in Fig.~\ref{fig:limit}. 
We note that the S/N is determined at the peak of the emission signal (namely the amplitude of the box profile). Therefore, detecting a broader signal requires a lower level of the peak contrast, as shown in the black line in Fig.~\ref{fig:limit}. However, this does not mean that the detectability increases when it comes to broad signals. 
As the flux is distributed into wider velocity space due to broadening mechanisms, it actually requires larger amount of integrated flux (namely higher Equivalent Width) to detect broader signals. The corresponding Equivalent Width limit required for 5$\sigma$ detection is plotted in the red curve in Fig.~\ref{fig:limit}.

The width of the signal is affecting the S/N of a potential detection in the following ways. 
First, the level of white noise decreases with the square root of the line width. Although the residual noise in our case is not fully uncorrelated due to imperfect stellar and telluric corrections, the S/N can still be enhanced, if not as much as by a factor of $\sqrt{W}$.  
Secondly, for a larger $W$ a wider part of the planet emission line trail is masked out, changing the residual noise pattern of the data, as shown in Fig. ~\ref{fig:res1d}, where the observed residuals (solid black curves) for different widths are not identical. 
Furthermore, the detection of broad planetary signals is hindered by the self-subtraction problem as discussed in section~\ref{sec:v1}. Hence, the detectability drops significantly for widths larger than 25 \kms, which is comparable to the radial velocity change of the planet during a night.
Overall, the detection limit drops significantly towards larger $W$, but reaches a plateau at $W>20$ \kms, for which we reach a 5$\sigma$ limit of $4\times10^{-4}$.

\subsection{Validation of signal recovery}
\begin{figure}
   \centering
   \includegraphics[width=\hsize]{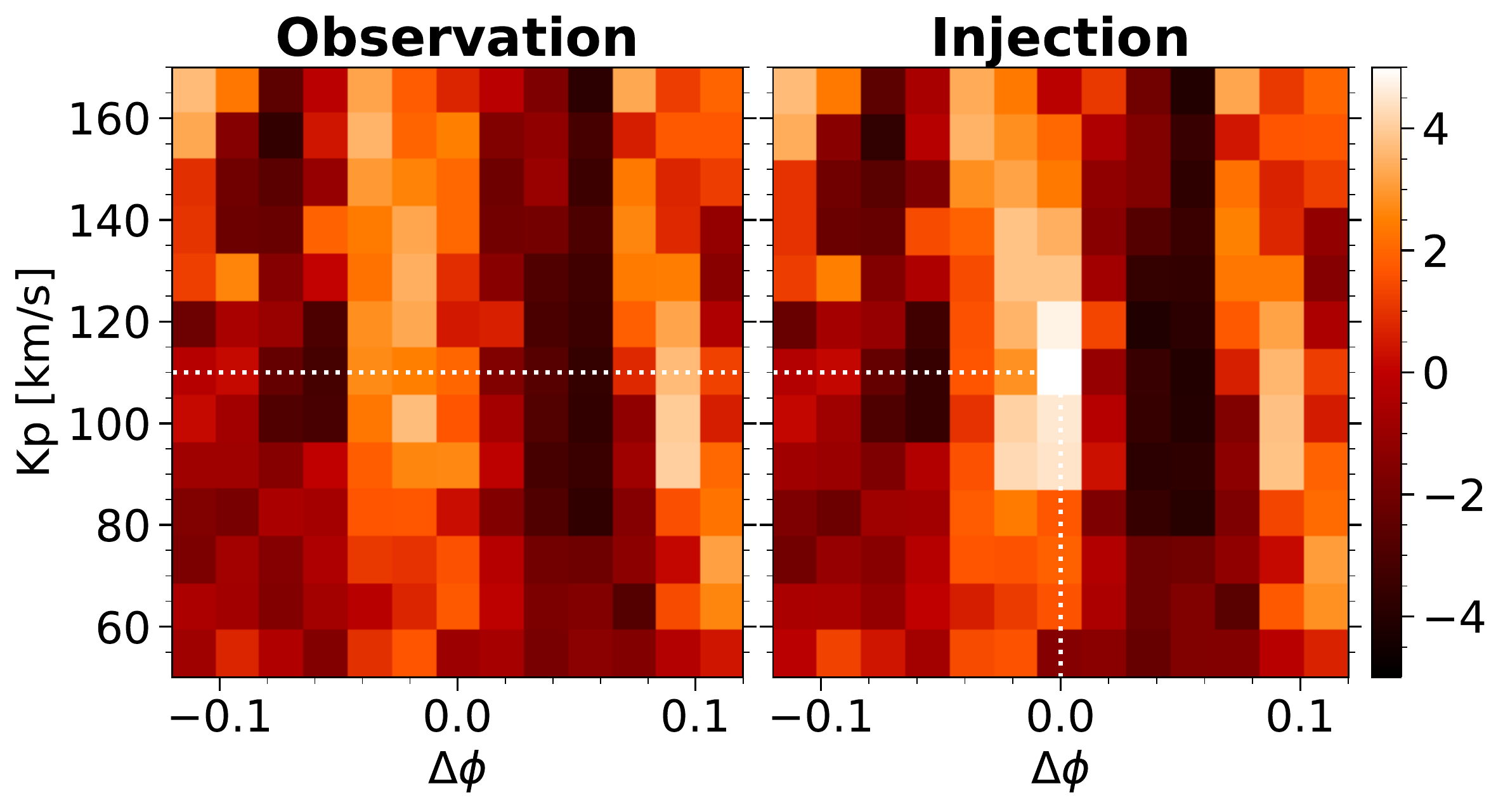}
   \caption{The S/N map for a grid of planetary orbits with different semi-amplitude $K_P$ and phase offset $\Delta\phi$. There is no salient signal in the observations (left panel). An injected signal at $K_P=110$ \kms and $\Delta\phi=0$ is clearly recovered (right panel).}
    \label{fig:map}%
\end{figure}

Prior knowledge of the planetary orbit was used to mask the trail in the spectral series to avoid self-subtraction of the potential signal (see section~\ref{sec:v1}). 
 In order to assess the influence of this procedure on the final result, we repeated the whole analysis adopting a grid of planetary orbits with different semi-amplitude $K_P$ and phase offset $\Delta\phi$ when applying the mask. In Fig.~\ref{fig:map}, we mapped the S/N of the residuals at 10830.3 \AA\, in the planetary rest frame for each hypothetical orbit in the grid. 
As a comparison, we produced a similar map when injecting an artificial planetary signal with a strength of $3.3\times10^{-4}$ and a width of 30 \kms, which is recovered at $K_P=110$ \kms and $\Delta\phi=0$ in Fig.~\ref{fig:map}, also without prior knowledge of the planetary orbit.

\section{Discussion}\label{sec:discuss}

\subsection{Detectability of He emission} \label{sec:theory_disc}

Using three nights of observations ($\sim$6.5 hours of integration) on \tauboo b using CARMENES, we reached a contrast limit of $4\times10^{-4}$ for a planetary \he emission with a boxcar profile with a width $>$ 20 \kms. However, following Equation~\ref{eq:1}, the estimated amplitude of the \he emission is typically $\leq 10^{-4}$. At this stage, no detection is expected with these observations. 

We note that we made several assumptions for simplicity. First, we assume the atmosphere extends significantly beyond the planet radius and the \he cloud is considered to be optically thin. The stellar radiation is absorbed uniformly in the cloud and reemitted isotropically. In addition, Equation~\ref{eq:1} assumes that the energy is conserved at the particular wavelength 10830 \AA, meaning that we only take into account the transition between the helium $2^3$S and $2^3$P states. 

In practice, the flux ratio can deviate from this simplified model in the following ways. (a) The flux emission of the cloud can be higher than our estimation if the extent of the exosphere is even larger than the stellar disk, in which case the measured $f_\mathrm{abs}$ in the transmission spectrum is lower than the actual absorption level of the entire cloud. 
(b) Analysis on the line ratio of \he triplet detected in transmission observations suggests that the planetary signal could also originate from an optically thick part of the atmosphere, especially for Jupiter-mass planets such as HD 189733b and HD 209458b, where the \he absorption feature traces compact atmospheres with an extent of only a fraction of planetary radii \citep{2018A&A...620A..97S, 2020A&A...636A..13L}.
If the cloud is not optically thin or the \he atmosphere is not significantly larger with respect to the planet, the reradiation is not isotropically emitted from the cloud in $4\pi$ but with preferential directions, as an analogy to scattered light. In this case, the emission flux at the full phase (i.e. when the illuminated hemisphere is totally in view) is enhanced as compared to the value in Equation~\ref{eq:1}, yet it also needs to be scaled with the phase function $\Phi_{\alpha}$ at different orbital phases (where we observe different portions of the illuminated hemisphere). (c) If the \he cloud is not optically thin, the planetary thermal emission passing through the cloud can be absorbed along the line of sight, which cancels out a portion of the airglow emission. To evaluate this effect, we compare the absorbed thermal emission to the airglow emission flux as follows. Assuming the planet emits as a blackbody with an equilibrium temperature of $T_p$, we can calculate the planetary thermal emission flux in contrast to the stellar flux at 10830 \AA. Taking \tauboo system as an example, the flux contrast is $\mathcal{F}_{th} / \mathcal{F}_\ast \sim 10^{-5} \sim 0.25 \mathcal{F}_{c} / \mathcal{F}_\ast$. To the optically thick limit, where all thermal emission is absorbed by the \he cloud, it counterbalances $\sim$25\% of the airglow emission. Consequently, even in the extreme optically thick case, the helium is expected to be seen in emission.
(d) The emission level of \he 10830 \AA\, is determined by the population of helium in the $2^3$P state as well as the rates of the stimulated and spontaneous decay of $2^3$P state. Studying this requires detailed accounts and comparisons of other de-populating processes including photoionization from $2^3$P state, electron/H-atom collisions, and radiative excitation to $2^3$D state (D3 line). Previous models of helium atmospheres, such as \citet{2018ApJ...855L..11O, 2020A&A...636A..13L}, neglect transitions related to $2^3$P state because the metastable $2^3$S state population is not significantly affected by these processes as an atom in $2^3$P state just decays back to $2^3$S state \citep{2018ApJ...855L..11O}. This indicates that the $2^3$P to $2^3$S decay is the major way of de-populating the $2^3$P state. 
Therefore, we can argue for the connection between the level of absorption and airglow emission at 10830 \AA.

In addition to the amplitude of the \he emission, the velocity profile is rather uncertain. 
Current detections suggest that the \he excess absorption spans a wide range of line widths. For instance, Saturn or Neptune mass planets such as WASP-69b and HAT-P-11b \citep{2018Sci...362.1384A, 2018Sci...362.1388N} are reported to have broad absorption feature (up to $\sim$30 \kms), while Jupiter-mass planets such as HD 209485b \citep{2019A&A...629A.110A} show a more narrow signature of $\sim$10 \kms, implying a lower rate of atmospheric escape and mass loss.
The line profile depends on the physical and hydrodynamic properties such as the kinematic temperature of the exosphere and the atmospheric escape, which are not well understood \citep{2016A&A...586A..75S}. 
Considering the diverse nature of planetary atmospheres, it is therefore recommended to explore a variety of line widths and line positions in data analysis.

\subsection{Future prospects}

\subsubsection{Observation design}

Based on our data analysis, we find several aspects that should be taken into account when designing future observations to search for \he airglow emission.
Firstly, the observations should avoid blending the planetary helium signal with strong telluric absorption lines (such as during our night 2 observations). In such situation, a large portion of the emission signal is removed along with the telluric correction. Even if the signal is preserved with other methods, strong telluric features cannot be corrected perfectly, introducing artifacts around the helium signal. Furthermore, the S/N of observations at the core of strong telluric absorption lines is significantly lower than in the continuum. Hence, the potential signal falling in such region is more difficult to detect.
Therefore, taking into account the relative position between the planetary and telluric lines, and selecting the proper time for observations according to the barycentric Earth radial velocity (BERV) and the planetary radial velocity are desirable.

The stellar activity is yet another factor to consider in data analysis because the stellar \he line as a chromospheric diagnostic may undergo temporal variation during observations. For an active star like \tauboo, this is likely to happen and introduce extra noise. 
\citet{2018AJ....156..189C}, \citet{2018A&A...620A..97S} and \citet{2020arXiv200505676G} evaluated the impact of stellar activity on probing the planetary helium atmosphere in transmission. However, unlike transit observations, additional noise related to stellar activity should not significantly affect  measurements of emission signals from a planet, because there is always a radial velocity offset between the star and planet when probing emission. Only when the radial velocity difference is small (that is, during transit and secondary eclipse), can stellar activity contribute to noise at the planetary rest frame and result in pseudo-signals. In terms of detecting airglow emission, the likely consequence of the variability of stellar \he line is to introduce systematic noise structure nearby the planetary signal in the temporal-combined residual spectrum, instead of affecting the level of planetary signal. Consequently, for highly active stars, there needs to be sufficient difference between the planet and stellar radial velocity, which depends on the orbital phase, during observations. 

\begin{figure}
   \centering
   \includegraphics[width=\hsize]{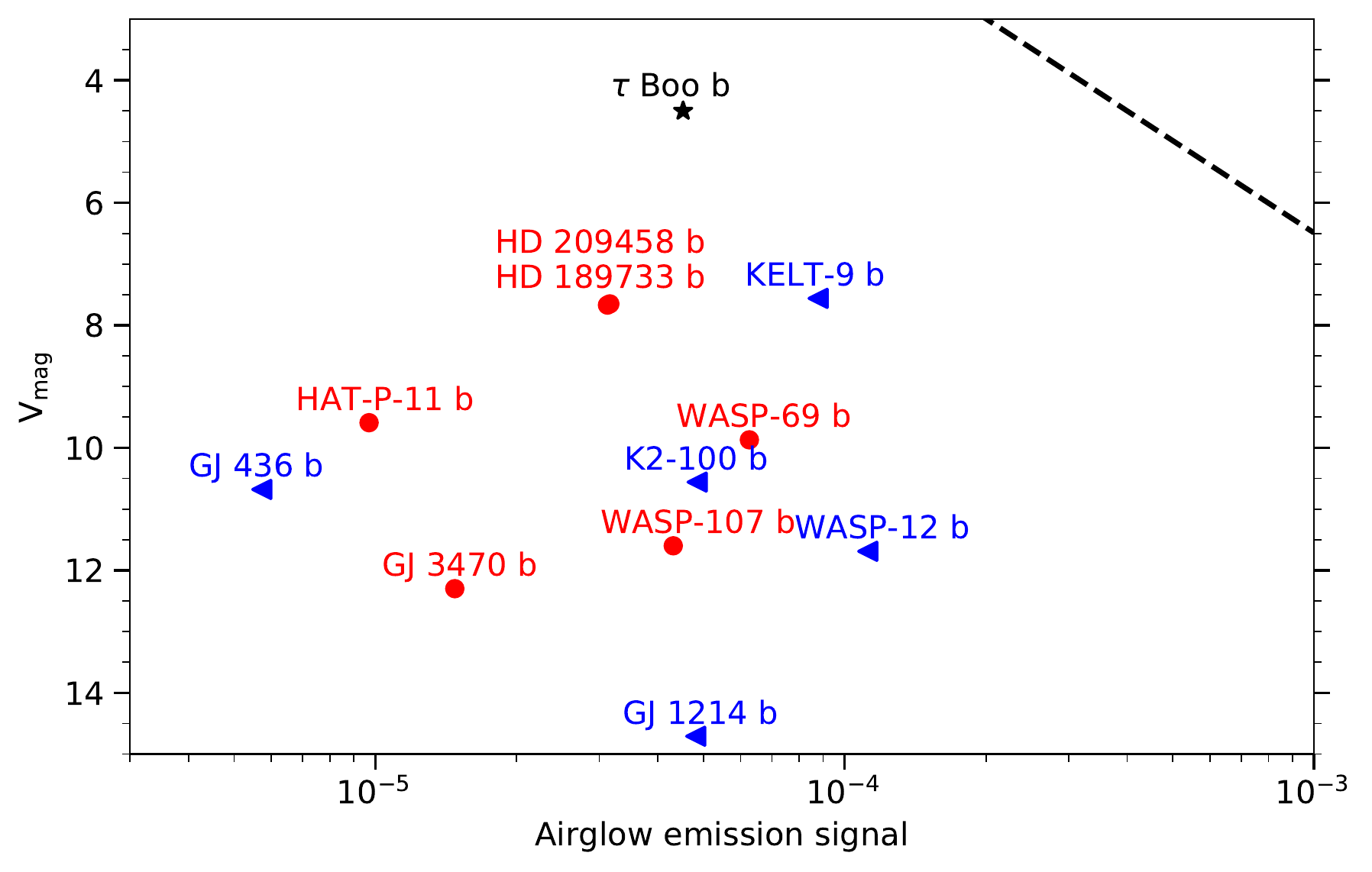}
   \caption{The V magnitude of host stars plotted against the airglow emission signal estimated using Equation~\ref{eq:1} for exoplanets with \he absorption detections (in red dots) or upper limits (in blue triangles). The black star represents the estimated signal of \tauboo b assuming a typical absorption level of 1\%. The black dashed line shows the detection limit of our analysis.}
    \label{fig:target}
\end{figure}

In terms of target selection, we performed the case study on \tauboo because the star is significantly brighter than other known hot-Jupiter systems, enabling high S/N observations. In addition, its high level of stellar activity likely results in an extended planetary atmosphere with the \he 2$^3$S populated.
As shown in Fig~\ref{fig:target}, accounting for both the stellar brightness and the airglow emission signal estimated using Equation~\ref{eq:1}, \tauboo b falls closest to our detection limit (denoted by the dashed line) among other potential targets. In the emission signal calculation, as we have no measurements of the \he absorption by the non-transiting \tauboo b, we assume an absorption level of 1\%, which, however, brings about some uncertainty.
\tauboo as an F-type star may also have high levels of mid-UV radiation that can de-populate the helium triplet states, leading to a weaker signal. It remains unclear to what extent these various factors contribute to the \he signal as a whole. In addition, the planet mass is significantly higher than that of most hot Jupiters, increasing the surface gravity, possibly making atmospheric escape more difficult \citep{2016A&A...586A..75S}. 
A more secure choice is to investigate those targets with \he absorption already measured via transmission spectroscopy (denoted with red dots in Fig~\ref{fig:target}), so that we have an expectation of the emission amplitude beforehand. 
Another benefit of targeting a transit planet is that we could obtain the stellar spectrum without any contribution from the planet during secondary eclipse. Using that as a reference spectrum, we can avoid self-subtracting the planetary signal in analysis, which is particularly important for broad signatures. 
Furthermore, as discussed in Section~\ref{sec:theory_disc}, the airglow emission signal is partially counterbalanced by  absorption of planet thermal emission through the \he cloud when it is compact. Consequently, planets with puffy and optically thin \he clouds are better targets for this purpose.

\subsubsection{Search with VLT/CRIRES+ and E-ELT/HIRES}

CRIRES+ is an upgrade of the cryogenic high-resolution infrared echelle spectrograph (CRIRES) located at the focus of UT1 of the Very Large Telescope (VLT), providing a spectral resolving power of 100,000 \citep{2014SPIE.9147E..19F}. 
If we simply scale the expected S/N for observations with CRIRES+ by the larger telescope diameter of the 8.2m VLT compared to the 3.5m Calar Alto Telescope, also considering that the throughput of CRIRES+ may be a factor of 1.5 lower than that of CARMENES, while the observing condition at VLT site is generally better, we expect in the same observing time to go a factor $\sim$2 deeper. 
Accordingly, we expect to reach a detection limit of $2\times10^{-4}$ for a broad \he emission given the same amount of integration ($\sim$6.5 hours). Taking \tauboo b as an example, following Equation~\ref{eq:1}, we estimate the amplitude of the emission to be $\sim5\times10^{-5}$ assuming a typical "transit" absorption level of 1\%. This means that a $3\sigma$ detection of \he airglow emission could be obtained in $\sim$5 nights of observations. 
Although probing \he emission is demanding using the current generation of telescopes, it is promising with the High Resolution Spectrograph (HIRES) at ESO’s forthcoming Extremely Large Telescope (E-ELT) \citep{2016SPIE.9908E..23M}. Thanks to the significantly larger diameter of the 39m E-ELT, the $5\sigma$ detection of \he airglow emission could be achieved with $\sim$3 hours of integration.
Of course there is a large uncertainty in the expected level of emission, which can be at least partly mitigated by choosing a transiting planet. 


\section{Conclusions}
This paper explores the possibility of probing \he emission at 10830 \AA\, in the extended atmospheres of hot Jupiters, the amplitude of which is estimated based on the \he absorption level as observed during transit. 
We search for this emission from \tauboo b using CARMENES spectra. We correct for the telluric and stellar features nearby the expected \he signal and combine multiple nights of observations. Given our estimation of the contrast of $\leq 10^{-4}$, the 6.5-hour data combined is not sufficient to put meaningful constrains on the \he emission from the planet. The detection limit that we derive from our analysis is $4\times 10^{-4}$ for a broad emission signal with a width of $>$20 \kms. While we do not reach the required contrast with the current CARMENES data, the \he airglow emission from hot-Jupiters is still promising to probe with future instruments. 

\begin{acknowledgements}
We thank the referee and editor for comments which helped improving the quality of the paper. Y.Z., I.S., F.A., and P.M. acknowledge funding from the European Research Council (ERC) under the European Union's Horizon 2020 research and innovation program under grant agreement No 694513.
P.M. acknowledges support from the European Research Council under the European Union’s Horizon 2020 research and innovation program under grant agreement No. 832428.
M.B. acknowledges support from the UK Science and Technology Facilities Council (STFC) research grant ST/S000631/1.
A.W. acknowledges the financial support of the SNSF by grant number P2GEP2-178191 and P400P2-186765.
      
\end{acknowledgements}

\bibliographystyle{aa} 
\bibliography{ref}

\begin{thebibliography}{42}
\expandafter\ifx\csname natexlab\endcsname\relax\def\natexlab#1{#1}\fi

\bibitem[{{Allart} {et~al.}(2019){Allart}, {Bourrier}, {Lovis}, {Ehrenreich},
  {Aceituno}, {Guijarro}, {Pepe}, {Sing}, {Spake}, \&
  {Wyttenbach}}]{2019A&A...623A..58A}
{Allart}, R., {Bourrier}, V., {Lovis}, C., {et~al.} 2019, \aap, 623, A58

\bibitem[{{Allart} {et~al.}(2018){Allart}, {Bourrier}, {Lovis}, {Ehrenreich},
  {Spake}, {Wyttenbach}, {Pino}, {Pepe}, {Sing}, \& {Lecavelier des
  Etangs}}]{2018Sci...362.1384A}
{Allart}, R., {Bourrier}, V., {Lovis}, C., {et~al.} 2018, Science, 362, 1384

\bibitem[{{Alonso-Floriano} {et~al.}(2019){Alonso-Floriano}, {Snellen},
  {Czesla}, {Bauer}, {Salz}, {Lamp{\'o}n}, {Lara}, {Nagel},
  {L{\'o}pez-Puertas}, {Nortmann}, {S{\'a}nchez-L{\'o}pez}, {Sanz-Forcada},
  {Caballero}, {Reiners}, {Ribas}, {Quirrenbach}, {Amado}, {Aceituno},
  {Anglada-Escud{\'e}}, {B{\'e}jar}, {Brinkm{\"o}ller}, {Hatzes}, {Henning},
  {Kaminski}, {K{\"u}rster}, {Labarga}, {Montes}, {Pall{\'e}}, {Schmitt}, \&
  {Zapatero Osorio}}]{2019A&A...629A.110A}
{Alonso-Floriano}, F.~J., {Snellen}, I.~A.~G., {Czesla}, S., {et~al.} 2019,
  \aap, 629, A110

\bibitem[{{Andretta} {et~al.}(2017){Andretta}, {Giampapa}, {Covino}, {Reiners},
  \& {Beeck}}]{2017ApJ...839...97A}
{Andretta}, V., {Giampapa}, M.~S., {Covino}, E., {Reiners}, A., \& {Beeck}, B.
  2017, \apj, 839, 97

\bibitem[{{Borsa} {et~al.}(2015){Borsa}, {Scandariato}, {Rainer}, {Bignamini},
  {Maggio}, {Poretti}, {Lanza}, {Di Mauro}, {Benatti}, {Biazzo}, {Bonomo},
  {Damasso}, {Esposito}, {Gratton}, {Affer}, {Barbieri}, {Boccato}, {Claudi},
  {Cosentino}, {Covino}, {Desidera}, {Fiorenzano}, {Gandolfi}, {Harutyunyan},
  {Maldonado}, {Micela}, {Molaro}, {Molinari}, {Pagano}, {Pillitteri},
  {Piotto}, {Shkolnik}, {Silvotti}, {Smareglia}, {Southworth}, {Sozzetti}, \&
  {Stelzer}}]{2015A&A...578A..64B}
{Borsa}, F., {Scandariato}, G., {Rainer}, M., {et~al.} 2015, \aap, 578, A64

\bibitem[{{Bourrier} {et~al.}(2018){Bourrier}, {Lecavelier des Etangs},
  {Ehrenreich}, {Sanz-Forcada}, {Allart}, {Ballester}, {Buchhave}, {Cohen},
  {Deming}, {Evans}, {Garc{\'\i}a Mu{\~n}oz}, {Henry}, {Kataria}, {Lavvas},
  {Lewis}, {L{\'o}pez-Morales}, {Marley}, {Sing}, \&
  {Wakeford}}]{2018A&A...620A.147B}
{Bourrier}, V., {Lecavelier des Etangs}, A., {Ehrenreich}, D., {et~al.} 2018,
  \aap, 620, A147

\bibitem[{{Brogi} {et~al.}(2012){Brogi}, {Snellen}, {de Kok}, {Albrecht},
  {Birkby}, \& {de Mooij}}]{2012Natur.486..502B}
{Brogi}, M., {Snellen}, I. A.~G., {de Kok}, R.~J., {et~al.} 2012, \nat, 486,
  502

\bibitem[{{Butler} {et~al.}(1997){Butler}, {Marcy}, {Williams}, {Hauser}, \&
  {Shirts}}]{1997ApJ...474L.115B}
{Butler}, R.~P., {Marcy}, G.~W., {Williams}, E., {Hauser}, H., \& {Shirts}, P.
  1997, \apjl, 474, L115

\bibitem[{{Caballero} {et~al.}(2016){Caballero}, {Gu{\`a}rdia}, {L{\'o}pez del
  Fresno}, {Zechmeister}, {de Juan}, {Alonso-Floriano}, {Amado}, {Colom{\'e}},
  {Cort{\'e}s-Contreras}, {Garc{\'\i}a-Piquer}, {Gesa}, {de Guindos}, {Hagen},
  {Helmling}, {Hern{\'a}ndez Casta{\~n}o}, {K{\"u}rster}, {L{\'o}pez-Santiago},
  {Montes}, {Morales Mu{\~n}oz}, {Pavlov}, {Quirrenbach}, {Reiners}, {Ribas},
  {Seifert}, \& {Solano}}]{2016SPIE.9910E..0EC}
{Caballero}, J.~A., {Gu{\`a}rdia}, J., {L{\'o}pez del Fresno}, M., {et~al.}
  2016, in Society of Photo-Optical Instrumentation Engineers (SPIE) Conference
  Series, Vol. 9910, \procspie, 99100E

\bibitem[{{Cauley} {et~al.}(2018){Cauley}, {Kuckein}, {Redfield}, {Shkolnik},
  {Denker}, {Llama}, \& {Verma}}]{2018AJ....156..189C}
{Cauley}, P.~W., {Kuckein}, C., {Redfield}, S., {et~al.} 2018, \aj, 156, 189

\bibitem[{{Crossfield} {et~al.}(2019){Crossfield}, {Barman}, {Hansen}, \&
  {Frewen}}]{2019RNAAS...3...24C}
{Crossfield}, I.~J.~M., {Barman}, T., {Hansen}, B., \& {Frewen}, S. 2019,
  Research Notes of the American Astronomical Society, 3, 24

\bibitem[{{Ehrenreich} {et~al.}(2015){Ehrenreich}, {Bourrier}, {Wheatley},
  {Lecavelier des Etangs}, {H{\'e}brard}, {Udry}, {Bonfils}, {Delfosse},
  {D{\'e}sert}, {Sing}, \& {Vidal-Madjar}}]{2015Natur.522..459E}
{Ehrenreich}, D., {Bourrier}, V., {Wheatley}, P.~J., {et~al.} 2015, \nat, 522,
  459

\bibitem[{{Follert} {et~al.}(2014){Follert}, {Dorn}, {Oliva}, {Lizon},
  {Hatzes}, {Piskunov}, {Reiners}, {Seemann}, {Stempels}, {Heiter}, {Marquart},
  {Lockhart}, {Anglada-Escude}, {L{\"o}winger}, {Baade}, {Grunhut}, {Bristow},
  {Klein}, {Jung}, {Ives}, {Kerber}, {Pozna}, {Paufique}, {Kaeufl}, {Origlia},
  {Valenti}, {Gojak}, {Hilker}, {Pasquini}, {Smette}, \&
  {Smoker}}]{2014SPIE.9147E..19F}
{Follert}, R., {Dorn}, R.~J., {Oliva}, E., {et~al.} 2014, in Society of
  Photo-Optical Instrumentation Engineers (SPIE) Conference Series, Vol. 9147,
  \procspie, 914719

\bibitem[{{Gaia Collaboration} {et~al.}(2018){Gaia Collaboration}, {Brown},
  {Vallenari}, {Prusti}, {de Bruijne}, {Babusiaux}, {Bailer-Jones}, {Biermann},
  {Evans}, {Eyer}, {Jansen}, {Jordi}, {Klioner}, {Lammers}, {Lindegren},
  {Luri}, {Mignard}, {Panem}, {Pourbaix}, {Randich}, {Sartoretti}, {Siddiqui},
  {Soubiran}, {van Leeuwen}, {Walton}, {Arenou}, {Bastian}, {Cropper},
  {Drimmel}, {Katz}, {Lattanzi}, {Bakker}, {Cacciari}, {Casta{\~n}eda},
  {Chaoul}, {Cheek}, {De Angeli}, {Fabricius}, {Guerra}, {Holl}, {Masana},
  {Messineo}, {Mowlavi}, {Nienartowicz}, {Panuzzo}, {Portell}, {Riello},
  {Seabroke}, {Tanga}, {Th{\'e}venin}, {Gracia-Abril}, {Comoretto},
  {Garcia-Reinaldos}, {Teyssier}, {Altmann}, {Andrae}, {Audard},
  {Bellas-Velidis}, {Benson}, {Berthier}, {Blomme}, {Burgess}, {Busso},
  {Carry}, {Cellino}, {Clementini}, {Clotet}, {Creevey}, {Davidson}, {De
  Ridder}, {Delchambre}, {Dell'Oro}, {Ducourant},
  {Fern{\'a}ndez-Hern{\'a}ndez}, {Fouesneau}, {Fr{\'e}mat}, {Galluccio},
  {Garc{\'\i}a-Torres}, {Gonz{\'a}lez-N{\'u}{\~n}ez}, {Gonz{\'a}lez-Vidal},
  {Gosset}, {Guy}, {Halbwachs}, {Hambly}, {Harrison}, {Hern{\'a}ndez},
  {Hestroffer}, {Hodgkin}, {Hutton}, {Jasniewicz}, {Jean-Antoine-Piccolo},
  {Jordan}, {Korn}, {Krone-Martins}, {Lanzafame}, {Lebzelter}, {L{\"o}ffler},
  {Manteiga}, {Marrese}, {Mart{\'\i}n-Fleitas}, {Moitinho}, {Mora}, {Muinonen},
  {Osinde}, {Pancino}, {Pauwels}, {Petit}, {Recio-Blanco}, {Richards},
  {Rimoldini}, {Robin}, {Sarro}, {Siopis}, {Smith}, {Sozzetti}, {S{\"u}veges},
  {Torra}, {van Reeven}, {Abbas}, {Abreu Aramburu}, {Accart}, {Aerts},
  {Altavilla}, {{\'A}lvarez}, {Alvarez}, {Alves}, {Anderson}, {Andrei},
  {Anglada Varela}, {Antiche}, {Antoja}, {Arcay}, {Astraatmadja}, {Bach},
  {Baker}, {Balaguer-N{\'u}{\~n}ez}, {Balm}, {Barache}, {Barata}, {Barbato},
  {Barblan}, {Barklem}, {Barrado}, {Barros}, {Barstow}, {Bartholom{\'e}
  Mu{\~n}oz}, {Bassilana}, {Becciani}, {Bellazzini}, {Berihuete}, {Bertone},
  {Bianchi}, {Bienaym{\'e}}, {Blanco-Cuaresma}, {Boch}, {Boeche}, {Bombrun},
  {Borrachero}, {Bossini}, {Bouquillon}, {Bourda}, {Bragaglia}, {Bramante},
  {Breddels}, {Bressan}, {Brouillet}, {Br{\"u}semeister}, {Brugaletta},
  {Bucciarelli}, {Burlacu}, {Busonero}, {Butkevich}, {Buzzi}, {Caffau},
  {Cancelliere}, {Cannizzaro}, {Cantat-Gaudin}, {Carballo}, {Carlucci},
  {Carrasco}, {Casamiquela}, {Castellani}, {Castro-Ginard}, {Charlot},
  {Chemin}, {Chiavassa}, {Cocozza}, {Costigan}, {Cowell}, {Crifo}, {Crosta},
  {Crowley}, {Cuypers}, {Dafonte}, {Damerdji}, {Dapergolas}, {David}, {David},
  {de Laverny}, {De Luise}, {De March}, {de Martino}, {de Souza}, {de Torres},
  {Debosscher}, {del Pozo}, {Delbo}, {Delgado}, {Delgado}, {Di Matteo},
  {Diakite}, {Diener}, {Distefano}, {Dolding}, {Drazinos}, {Dur{\'a}n},
  {Edvardsson}, {Enke}, {Eriksson}, {Esquej}, {Eynard Bontemps}, {Fabre},
  {Fabrizio}, {Faigler}, {Falc{\~a}o}, {Farr{\`a}s Casas}, {Federici},
  {Fedorets}, {Fernique}, {Figueras}, {Filippi}, {Findeisen}, {Fonti},
  {Fraile}, {Fraser}, {Fr{\'e}zouls}, {Gai}, {Galleti}, {Garabato},
  {Garc{\'\i}a-Sedano}, {Garofalo}, {Garralda}, {Gavel}, {Gavras}, {Gerssen},
  {Geyer}, {Giacobbe}, {Gilmore}, {Girona}, {Giuffrida}, {Glass}, {Gomes},
  {Granvik}, {Gueguen}, {Guerrier}, {Guiraud}, {Guti{\'e}rrez-S{\'a}nchez},
  {Haigron}, {Hatzidimitriou}, {Hauser}, {Haywood}, {Heiter}, {Helmi}, {Heu},
  {Hilger}, {Hobbs}, {Hofmann}, {Holland}, {Huckle}, {Hypki}, {Icardi},
  {Jan{\ss}en}, {Jevardat de Fombelle}, {Jonker}, {Juh{\'a}sz}, {Julbe},
  {Karampelas}, {Kewley}, {Klar}, {Kochoska}, {Kohley}, {Kolenberg},
  {Kontizas}, {Kontizas}, {Koposov}, {Kordopatis}, {Kostrzewa-Rutkowska},
  {Koubsky}, {Lambert}, {Lanza}, {Lasne}, {Lavigne}, {Le Fustec}, {Le
  Poncin-Lafitte}, {Lebreton}, {Leccia}, {Leclerc}, {Lecoeur-Taibi},
  {Lenhardt}, {Leroux}, {Liao}, {Licata}, {Lindstr{\o}m}, {Lister}, {Livanou},
  {Lobel}, {L{\'o}pez}, {Managau}, {Mann}, {Mantelet}, {Marchal}, {Marchant},
  {Marconi}, {Marinoni}, {Marschalk{\'o}}, {Marshall}, {Martino}, {Marton},
  {Mary}, {Massari}, {Matijevi{\v{c}}}, {Mazeh}, {McMillan}, {Messina},
  {Michalik}, {Millar}, {Molina}, {Molinaro}, {Moln{\'a}r}, {Montegriffo},
  {Mor}, {Morbidelli}, {Morel}, {Morris}, {Mulone}, {Muraveva}, {Musella},
  {Nelemans}, {Nicastro}, {Noval}, {O'Mullane}, {Ord{\'e}novic},
  {Ord{\'o}{\~n}ez-Blanco}, {Osborne}, {Pagani}, {Pagano}, {Pailler},
  {Palacin}, {Palaversa}, {Panahi}, {Pawlak}, {Piersimoni}, {Pineau}, {Plachy},
  {Plum}, {Poggio}, {Poujoulet}, {Pr{\v{s}}a}, {Pulone}, {Racero}, {Ragaini},
  {Rambaux}, {Ramos-Lerate}, {Regibo}, {Reyl{\'e}}, {Riclet}, {Ripepi}, {Riva},
  {Rivard}, {Rixon}, {Roegiers}, {Roelens}, {Romero-G{\'o}mez}, {Rowell},
  {Royer}, {Ruiz-Dern}, {Sadowski}, {Sagrist{\`a} Sell{\'e}s}, {Sahlmann},
  {Salgado}, {Salguero}, {Sanna}, {Santana-Ros}, {Sarasso}, {Savietto},
  {Schultheis}, {Sciacca}, {Segol}, {Segovia}, {S{\'e}gransan}, {Shih},
  {Siltala}, {Silva}, {Smart}, {Smith}, {Solano}, {Solitro}, {Sordo}, {Soria
  Nieto}, {Souchay}, {Spagna}, {Spoto}, {Stampa}, {Steele},
  {Steidelm{\"u}ller}, {Stephenson}, {Stoev}, {Suess}, {Surdej}, {Szabados},
  {Szegedi-Elek}, {Tapiador}, {Taris}, {Tauran}, {Taylor}, {Teixeira},
  {Terrett}, {Teyssand ier}, {Thuillot}, {Titarenko}, {Torra Clotet}, {Turon},
  {Ulla}, {Utrilla}, {Uzzi}, {Vaillant}, {Valentini}, {Valette}, {van Elteren},
  {Van Hemelryck}, {van Leeuwen}, {Vaschetto}, {Vecchiato}, {Veljanoski},
  {Viala}, {Vicente}, {Vogt}, {von Essen}, {Voss}, {Votruba}, {Voutsinas},
  {Walmsley}, {Weiler}, {Wertz}, {Wevers}, {Wyrzykowski}, {Yoldas},
  {{\v{Z}}erjal}, {Ziaeepour}, {Zorec}, {Zschocke}, {Zucker}, {Zurbach}, \&
  {Zwitter}}]{2018A&A...616A...1G}
{Gaia Collaboration}, {Brown}, A.~G.~A., {Vallenari}, A., {et~al.} 2018, \aap,
  616, A1

\bibitem[{{Gaidos} {et~al.}(2020){Gaidos}, {Hirano}, {Mann}, {Owens}, {Berger},
  {France}, {Vanderburg}, {Harakawa}, {Hodapp}, {Ishizuka}, {Jacobson},
  {Konishi}, {Kotani}, {Kudo}, {Kurokawa}, {Kuzuhara}, {Nishikawa}, {Omiya},
  {Serizawa}, {Tamura}, \& {Ueda}}]{2020MNRAS.495..650G}
{Gaidos}, E., {Hirano}, T., {Mann}, A.~W., {et~al.} 2020, \mnras, 495, 650

\bibitem[{{Guilluy} {et~al.}(2020){Guilluy}, {Andretta}, {Borsa}, {Giacobbe},
  {Sozzetti}, {Covino}, {Bourrier}, {Fossati}, {Bonomo}, {Esposito},
  {Giampapa}, {Harutyunyan}, {Rainer}, {Brogi}, {Bruno}, {Claudi}, {Frustagli},
  {Lanza}, {Mancini}, {Pino}, {Poretti}, {Scandariato}, {Affer}, {Baffa},
  {Baruffolo}, {Benatti}, {Biazzo}, {Bignamini}, {Boschin}, {Carleo},
  {Cecconi}, {Cosentino}, {Damasso}, {Desidera}, {Falcini}, {Martinez
  Fiorenzano}, {Ghedina}, {Gonz{\'a}lez-{\'A}lvarez}, {Guerra}, {Hernandez},
  {Leto}, {Maggio}, {Malavolta}, {Maldonado}, {Micela}, {Molinari},
  {Nascimbeni}, {Pagano}, {Pedani}, {Piotto}, \&
  {Reiners}}]{2020arXiv200505676G}
{Guilluy}, G., {Andretta}, V., {Borsa}, F., {et~al.} 2020, arXiv e-prints,
  arXiv:2005.05676

\bibitem[{{Hirano} {et~al.}(2020){Hirano}, {Krishnamurthy}, {Gaidos},
  {Flewelling}, {Mann}, {Narita}, {Plavchan}, {Kotani}, {Tamura}, {Harakawa},
  {Hodapp}, {Ishizuka}, {Jacobson}, {Konishi}, {Kudo}, {Kurokawa}, {Kuzuhara},
  {Nishikawa}, {Omiya}, {Serizawa}, {Ueda}, \& {Vievard}}]{2020arXiv200613243H}
{Hirano}, T., {Krishnamurthy}, V., {Gaidos}, E., {et~al.} 2020, arXiv e-prints,
  arXiv:2006.13243

\bibitem[{{Huensch} {et~al.}(1998){Huensch}, {Schmitt}, \&
  {Voges}}]{1998A&AS..132..155H}
{Huensch}, M., {Schmitt}, J.~H.~M.~M., \& {Voges}, W. 1998, \aaps, 132, 155

\bibitem[{{Justesen} \& {Albrecht}(2019)}]{2019A&A...625A..59J}
{Justesen}, A.~B. \& {Albrecht}, S. 2019, \aap, 625, A59

\bibitem[{{Kirk} {et~al.}(2020){Kirk}, {Alam}, {L{\'o}pez-Morales}, \&
  {Zeng}}]{2020AJ....159..115K}
{Kirk}, J., {Alam}, M.~K., {L{\'o}pez-Morales}, M., \& {Zeng}, L. 2020, \aj,
  159, 115

\bibitem[{{Kreidberg} \& {Oklop{\v{c}}i{\'c}}(2018)}]{2018RNAAS...2...44K}
{Kreidberg}, L. \& {Oklop{\v{c}}i{\'c}}, A. 2018, Research Notes of the
  American Astronomical Society, 2, 44

\bibitem[{{Kulow} {et~al.}(2014){Kulow}, {France}, {Linsky}, \&
  {Loyd}}]{2014ApJ...786..132K}
{Kulow}, J.~R., {France}, K., {Linsky}, J., \& {Loyd}, R.~O.~P. 2014, \apj,
  786, 132

\bibitem[{{Lamp{\'o}n} {et~al.}(2020){Lamp{\'o}n}, {L{\'o}pez-Puertas}, {Lara},
  {S{\'a}nchez-L{\'o}pez}, {Salz}, {Czesla}, {Sanz-Forcada}, {Molaverdikhani},
  {Alonso-Floriano}, {Nortmann}, {Caballero}, {Bauer}, {Pall{\'e}}, {Montes},
  {Quirrenbach}, {Nagel}, {Ribas}, {Reiners}, \& {Amado}}]{2020A&A...636A..13L}
{Lamp{\'o}n}, M., {L{\'o}pez-Puertas}, M., {Lara}, L.~M., {et~al.} 2020, \aap,
  636, A13

\bibitem[{{Lockwood} {et~al.}(2014){Lockwood}, {Johnson}, {Bender}, {Carr},
  {Barman}, {Richert}, \& {Blake}}]{2014ApJ...783L..29L}
{Lockwood}, A.~C., {Johnson}, J.~A., {Bender}, C.~F., {et~al.} 2014, \apjl,
  783, L29

\bibitem[{{Mansfield} {et~al.}(2018){Mansfield}, {Bean}, {Oklop{\v{c}}i{\'c}},
  {Kreidberg}, {D{\'e}sert}, {Kempton}, {Line}, {Fortney}, {Henry}, {Mallonn},
  {Stevenson}, {Dragomir}, {Allart}, \& {Bourrier}}]{2018ApJ...868L..34M}
{Mansfield}, M., {Bean}, J.~L., {Oklop{\v{c}}i{\'c}}, A., {et~al.} 2018, \apjl,
  868, L34

\bibitem[{{Marconi} {et~al.}(2016){Marconi}, {Di Marcantonio}, {D'Odorico},
  {Cristiani}, {Maiolino}, {Oliva}, {Origlia}, {Riva}, {Valenziano}, {Zerbi},
  {Abreu}, {Adibekyan}, {Allende Prieto}, {Amado}, {Benz}, {Boisse}, {Bonfils},
  {Bouchy}, {Buchhave}, {Buscher}, {Cabral}, {Canto Martins}, {Chiavassa},
  {Coelho}, {Christensen}, {Delgado-Mena}, {de Medeiros}, {Di Varano},
  {Figueira}, {Fisher}, {Fynbo}, {Glasse}, {Haehnelt}, {Haniff}, {Hansen},
  {Hatzes}, {Huke}, {Korn}, {Le{\~a}o}, {Liske}, {Lovis}, {Maslowski},
  {Matute}, {McCracken}, {Martins}, {Monteiro}, {Morris}, {Morris}, {Nicklas},
  {Niedzielski}, {Nunes}, {Palle}, {Parr-Burman}, {Parro}, {Parry}, {Pepe},
  {Piskunov}, {Queloz}, {Quirrenbach}, {Rebolo Lopez}, {Reiners}, {Reid},
  {Santos}, {Seifert}, {Sousa}, {Stempels}, {Strassmeier}, {Sun}, {Udry},
  {Vanzi}, {Vestergaard}, {Weber}, \& {Zackrisson}}]{2016SPIE.9908E..23M}
{Marconi}, A., {Di Marcantonio}, P., {D'Odorico}, V., {et~al.} 2016, in Society
  of Photo-Optical Instrumentation Engineers (SPIE) Conference Series, Vol.
  9908, \procspie, 990823

\bibitem[{{Ninan} {et~al.}(2020){Ninan}, {Stefansson}, {Mahadevan}, {Bender},
  {Robertson}, {Ramsey}, {Terrien}, {Wright}, {Diddams}, {Kanodia}, {Cochran},
  {Endl}, {Ford}, {Fredrick}, {Halverson}, {Hearty}, {Jennings}, {Kaplan},
  {Lubar}, {Metcalf}, {Monson}, {Nitroy}, {Roy}, \&
  {Schwab}}]{2020ApJ...894...97N}
{Ninan}, J.~P., {Stefansson}, G., {Mahadevan}, S., {et~al.} 2020, \apj, 894, 97

\bibitem[{{Nortmann} {et~al.}(2018){Nortmann}, {Pall{\'e}}, {Salz},
  {Sanz-Forcada}, {Nagel}, {Alonso-Floriano}, {Czesla}, {Yan}, {Chen},
  {Snellen}, {Zechmeister}, {Schmitt}, {L{\'o}pez-Puertas}, {Casasayas-Barris},
  {Bauer}, {Amado}, {Caballero}, {Dreizler}, {Henning}, {Lamp{\'o}n}, {Montes},
  {Molaverdikhani}, {Quirrenbach}, {Reiners}, {Ribas}, {S{\'a}nchez-L{\'o}pez},
  {Schneider}, \& {Zapatero Osorio}}]{2018Sci...362.1388N}
{Nortmann}, L., {Pall{\'e}}, E., {Salz}, M., {et~al.} 2018, Science, 362, 1388

\bibitem[{{Oklop{\v{c}}i{\'c}}(2019)}]{2019ApJ...881..133O}
{Oklop{\v{c}}i{\'c}}, A. 2019, \apj, 881, 133

\bibitem[{{Oklop{\v{c}}i{\'c}} \& {Hirata}(2018)}]{2018ApJ...855L..11O}
{Oklop{\v{c}}i{\'c}}, A. \& {Hirata}, C.~M. 2018, \apjl, 855, L11

\bibitem[{{Owen}(2019)}]{2019AREPS..47...67O}
{Owen}, J.~E. 2019, Annual Review of Earth and Planetary Sciences, 47, 67

\bibitem[{{Palle} {et~al.}(2020){Palle}, {Nortmann}, {Casasayas-Barris},
  {Lamp{\'o}n}, {L{\'o}pez-Puertas}, {Caballero}, {Sanz-Forcada}, {Lara},
  {Nagel}, {Yan}, {Alonso-Floriano}, {Amado}, {Chen}, {Cifuentes},
  {Cort{\'e}s-Contreras}, {Czesla}, {Molaverdikhani}, {Montes}, {Passegger},
  {Quirrenbach}, {Reiners}, {Ribas}, {S{\'a}nchez-L{\'o}pez}, {Schweitzer},
  {Stangret}, {Zapatero Osorio}, \& {Zechmeister}}]{2020A&A...638A..61P}
{Palle}, E., {Nortmann}, L., {Casasayas-Barris}, N., {et~al.} 2020, \aap, 638,
  A61

\bibitem[{{Quirrenbach} {et~al.}(2016){Quirrenbach}, {Amado}, {Caballero},
  {Mundt}, {Reiners}, {Ribas}, {Seifert}, {Abril}, {Aceituno},
  {Alonso-Floriano}, {Anwand -Heerwart}, {Azzaro}, {Bauer}, {Barrado},
  {Becerril}, {Bejar}, {Benitez}, {Berdinas}, {Brinkm{\"o}ller}, {Cardenas},
  {Casal}, {Claret}, {Colom{\'e}}, {Cortes-Contreras}, {Czesla}, {Doellinger},
  {Dreizler}, {Feiz}, {Fernandez}, {Ferro}, {Fuhrmeister}, {Galadi},
  {Gallardo}, {G{\'a}lvez-Ortiz}, {Garcia-Piquer}, {Garrido}, {Gesa},
  {G{\'o}mez Galera}, {Gonz{\'a}lez Hern{\'a}ndez}, {Gonzalez Peinado},
  {Gr{\"o}zinger}, {Gu{\`a}rdia}, {Guenther}, {de Guindos}, {Hagen}, {Hatzes},
  {Hauschildt}, {Helmling}, {Henning}, {Hermann}, {Hern{\'a}ndez Arabi},
  {Hern{\'a}ndez Casta{\~n}o}, {Hern{\'a}ndez Hernando}, {Herrero}, {Huber},
  {Huber}, {Huke}, {Jeffers}, {de Juan}, {Kaminski}, {Kehr}, {Kim}, {Klein},
  {Kl{\"u}ter}, {K{\"u}rster}, {Lafarga}, {Lara}, {Lamert}, {Laun},
  {Launhardt}, {Lemke}, {Lenzen}, {Llamas}, {Lopez del Fresno},
  {L{\'o}pez-Puertas}, {L{\'o}pez-Santiago}, {Lopez Salas}, {Magan
  Madinabeitia}, {Mall}, {Mandel}, {Mancini}, {Marin Molina}, {Maroto
  Fern{\'a}ndez}, {Mart{\'\i}n}, {Mart{\'\i}n-Ruiz}, {Marvin}, {Mathar},
  {Mirabet}, {Montes}, {Morales}, {Morales Mu{\~n}oz}, {Nagel}, {Naranjo},
  {Nowak}, {Palle}, {Panduro}, {Passegger}, {Pavlov}, {Pedraz}, {Perez},
  {P{\'e}rez-Medialdea}, {Perger}, {Pluto}, {Ram{\'o}n}, {Rebolo}, {Redondo},
  {Reffert}, {Reinhart}, {Rhode}, {Rix}, {Rodler}, {Rodr{\'\i}guez},
  {Rodr{\'\i}guez L{\'o}pez}, {Rohloff}, {Rosich}, {Sanchez Carrasco},
  {Sanz-Forcada}, {Sarkis}, {Sarmiento}, {Sch{\"a}fer}, {Schiller}, {Schmidt},
  {Schmitt}, {Sch{\"o}fer}, {Schweitzer}, {Shulyak}, {Solano}, {Stahl},
  {Storz}, {Tabernero}, {Tala}, {Tal-Or}, {Ulbrich}, {Veredas}, {Vico Linares},
  {Vilardell}, {Wagner}, {Winkler}, {Zapatero Osorio}, {Zechmeister},
  {Ammler-von Eiff}, {Anglada-Escud{\'e}}, {del Burgo}, {Garcia-Vargas},
  {Klutsch}, {Lizon}, {Lopez-Morales}, {Ofir}, {P{\'e}rez-Calpena}, {Perryman},
  {S{\'a}nchez-Blanco}, {Strachan}, {St{\"u}rmer}, {Su{\'a}rez}, {Trifonov},
  {Tulloch}, \& {Xu}}]{2016SPIE.9908E..12Q}
{Quirrenbach}, A., {Amado}, P.~J., {Caballero}, J.~A., {et~al.} 2016, in
  Society of Photo-Optical Instrumentation Engineers (SPIE) Conference Series,
  Vol. 9908, \procspie, 990812

\bibitem[{{Salz} {et~al.}(2018){Salz}, {Czesla}, {Schneider}, {Nagel},
  {Schmitt}, {Nortmann}, {Alonso-Floriano}, {L{\'o}pez-Puertas}, {Lamp{\'o}n},
  {Bauer}, {Snellen}, {Pall{\'e}}, {Caballero}, {Yan}, {Chen}, {Sanz-Forcada},
  {Amado}, {Quirrenbach}, {Ribas}, {Reiners}, {B{\'e}jar}, {Casasayas-Barris},
  {Cort{\'e}s-Contreras}, {Dreizler}, {Guenther}, {Henning}, {Jeffers},
  {Kaminski}, {K{\"u}rster}, {Lafarga}, {Lara}, {Molaverdikhani}, {Montes},
  {Morales}, {S{\'a}nchez-L{\'o}pez}, {Seifert}, {Zapatero Osorio}, \&
  {Zechmeister}}]{2018A&A...620A..97S}
{Salz}, M., {Czesla}, S., {Schneider}, P.~C., {et~al.} 2018, \aap, 620, A97

\bibitem[{{Salz} {et~al.}(2016){Salz}, {Czesla}, {Schneider}, \&
  {Schmitt}}]{2016A&A...586A..75S}
{Salz}, M., {Czesla}, S., {Schneider}, P.~C., \& {Schmitt}, J.~H.~M.~M. 2016,
  \aap, 586, A75

\bibitem[{{Sanz-Forcada} {et~al.}(2011){Sanz-Forcada}, {Micela}, {Ribas},
  {Pollock}, {Eiroa}, {Velasco}, {Solano}, \&
  {Garc{\'\i}a-{\'A}lvarez}}]{2011A&A...532A...6S}
{Sanz-Forcada}, J., {Micela}, G., {Ribas}, I., {et~al.} 2011, \aap, 532, A6

\bibitem[{{Seager}(2010)}]{2010eapp.book.....S}
{Seager}, S. 2010, {Exoplanet Atmospheres: Physical Processes}

\bibitem[{{Seager} \& {Sasselov}(2000)}]{2000ApJ...537..916S}
{Seager}, S. \& {Sasselov}, D.~D. 2000, \apj, 537, 916

\bibitem[{{Spake} {et~al.}(2018){Spake}, {Sing}, {Evans}, {Oklop{\v{c}}i{\'c}},
  {}, {Bourrier}, {Kreidberg}, {Rackham}, {Irwin}, {Ehrenreich}, {Wyttenbach},
  {Wakeford}, {Zhou}, {Chubb}, {Nikolov}, {Goyal}, {Henry}, {Williamson},
  {Blumenthal}, {Anderson}, {Hellier}, {Charbonneau}, {Udry}, \&
  {Madhusudhan}}]{2018Natur.557...68S}
{Spake}, J.~J., {Sing}, D.~K., {Evans}, T.~M., {et~al.} 2018, \nat, 557, 68

\bibitem[{{Takeda} {et~al.}(2007){Takeda}, {Ford}, {Sills}, {Rasio}, {Fischer},
  \& {Valenti}}]{2007ApJS..168..297T}
{Takeda}, G., {Ford}, E.~B., {Sills}, A., {et~al.} 2007, \apjs, 168, 297

\bibitem[{{Vidal-Madjar} {et~al.}(2003){Vidal-Madjar}, {Lecavelier des Etangs},
  {D{\'e}sert}, {Ballester}, {Ferlet}, {H{\'e}brard}, \&
  {Mayor}}]{2003Natur.422..143V}
{Vidal-Madjar}, A., {Lecavelier des Etangs}, A., {D{\'e}sert}, J.~M., {et~al.}
  2003, \nat, 422, 143

\bibitem[{{Wright} {et~al.}(2004){Wright}, {Marcy}, {Butler}, \&
  {Vogt}}]{2004ApJS..152..261W}
{Wright}, J.~T., {Marcy}, G.~W., {Butler}, R.~P., \& {Vogt}, S.~S. 2004, \apjs,
  152, 261

\end{thebibliography}

\begin{appendix} 
\section{Derivation of helium emission strength with radiative transfer}\label{app}

During transit, the stellar flux $\mathcal{F}_\ast$ decreases by $\Delta \mathcal{F}_\ast$ due to the absorption by helium atoms at 2$^3$S state. The absorption depth $T_\lambda$ measured by transmission spectroscopy is
\begin{equation}\label{eq:trans}
   T_\lambda =  \frac{\Delta \mathcal{F}_\ast}{\mathcal{F}_\ast} = \frac{R_c^2}{R_\ast^2} f_\mathrm{abs}= \frac{R_c^2}{R_\ast^2} (1-e^{-\tau_c}),
\end{equation}
where $\tau_c$ is the optical depth of the \he cloud around the planet, and the structure of the cloud is neglected. 

\subsection{Optically thick}
For optically thick clouds, namely, $\tau_c>>1$, Equation~\ref{eq:trans} becomes
\begin{equation}\label{eq:thick}
    R_c = R_\ast \sqrt{T_\lambda} \ .
\end{equation}
The emission flux from the cloud can be approximated as isotropic scattering with an albedo of 1, resulting in a factor of 2/3 of the received flux \citep[][Chap.~3]{2010eapp.book.....S}:
\begin{equation}\label{eq:f0}
    F_\mathrm{c}^S = \frac{2}{3} F_{\ast}^{S} \bigg(\frac{R_\ast}{a}\bigg)^2.
\end{equation}
Substituting Equation~\ref{eq:dillute} and \ref{eq:thick} into Equation~\ref{eq:f0}, we get
\begin{equation}
     \frac{\mathcal{F}_\mathrm{c}}{\mathcal{F}_\ast} = \frac{2}{3} \frac{\Delta \mathcal{F}_\ast}{\mathcal{F}_\ast} \bigg(\frac{R_\ast}{a}\bigg)^2 = \frac{2}{3} T_\lambda \bigg(\frac{R_\ast}{a}\bigg)^2.
\end{equation}

\subsection{Optically thin}

For optically thin cases, we consider the simplified 1D plane-parallel equation of radiative transfer
\begin{equation}
    \frac{dI}{d\tau} = I(\tau) - S(\tau),
\end{equation}
where $I_\nu$ is the spectral radiance, $S$ is the source function, and $\tau$ is the extinction optical depth which increases towards interior of the cloud (i.e. the optical depth at the interior is $\tau_c$ and that at the surface is 0).

\noindent The source function is approximated as
\begin{equation}
    S(\tau) = J_{\ast}(\tau) = \frac{1}{4\pi} \int I_{\ast}(\tau) d\Omega = \frac{1}{4\pi} \bigg(\frac{R_\ast}{a}\bigg)^2 F_{\ast}^{S} e^{-\tau}.
\end{equation}
Hence, the solution for the equation of radiative transfer can be composed as 
\begin{equation}
    I(\tau) = ce^{\tau}+\frac{1}{8\pi}\bigg(\frac{R_\ast}{a}\bigg)^2F_{\ast}^{S} e^{-\tau}.
\end{equation}
Using the boundary condition $I(\tau_c)=0$, the integral constant $c$ is determined, making the solution simplified as 
\begin{equation}
    I(\tau) = \frac{1}{8\pi}\bigg(\frac{R_\ast}{a}\bigg)^2F_{\ast}^{S} (e^{-\tau}-e^{\tau-2\tau_c}).
\end{equation}
Then the emergent radiance from the cloud $I_c$ is
\begin{equation}
    I_c = I(0) = \frac{1}{8\pi}\bigg(\frac{R_\ast}{a}\bigg)^2F_{\ast}^{S} (1-e^{-2\tau_c}).
\end{equation}
The observed flux of the cloud is 
\begin{equation}\label{eq:f1}
   \mathcal{F}_c = \int I_c \ (\boldsymbol{n}_{c} \cdot \boldsymbol{n}_\mathrm{detector}) \ d\Omega = \frac{1}{8\pi}\bigg(\frac{R_\ast}{a}\bigg)^2F_{\ast}^{S} (1-e^{-2\tau_c}) \frac{\pi R_c^2}{D^2},
\end{equation}
where $\boldsymbol{n}_{c}$ and $\boldsymbol{n}_\mathrm{detector}$ represent the unit vectors along the intensity of the cloud and the normal direction of the detector surface respectively. Since the targets are far away, we can safely assume $\boldsymbol{n}_{c} \cdot \boldsymbol{n}_\mathrm{detector} = 1$ in Equation~\ref{eq:f1}.

\noindent Substituting $\mathcal{F}_\ast = F_\ast^S R_\ast^2/D^2$ and Equation~\ref{eq:trans} into Equation~\ref{eq:f1}, we have
\begin{equation}\label{eq:f2}
     \mathcal{F}_{c} = \frac{1}{8} \Delta F_\ast \bigg(\frac{R_\ast}{a}\bigg)^2 \frac{1-e^{-2\tau_c}}{1-e^{-\tau_c}}.
\end{equation}
At the optically thin limit ($\tau_c<<1$), Equation~\ref{eq:f2} reduces to
\begin{equation}\label{eq:a}
     \frac{\mathcal{F}_{c}}{\mathcal{F}_\ast} = \frac{1}{4} \frac{\Delta \mathcal{F}_\ast}{\mathcal{F}_\ast} \bigg(\frac{R_\ast}{a}\bigg)^2 = \frac{1}{4} T_\lambda \bigg(\frac{R_\ast}{a}\bigg)^2,
\end{equation}
which agrees with Equation~\ref{eq:1}.

\end{appendix}

\end{document}